\documentstyle{l-aa}
\def\h{~$h^{-1}$ Mpc~}
\def\htre{~$h^{-3}$ Mpc$^3$~}
\def\hmentre{~$h^{3}$ Mpc$^{-3}$~}
\def\hmagp{$+5\log h$~}

\def\lsole{~L$_{\odot}$~}
\begin{document}
\thesaurus{ 11(11.04.1; 11.12.2) }
\title{ The ESO Slice Project (ESP) galaxy redshift survey:
\thanks{based on observations collected at the European Southern
Observatory, La Silla, Chile.} }
\subtitle{ II. The luminosity function and mean galaxy density}
%
%
\author{
E.Zucca\inst{1,2}
\and
G.Zamorani\inst{1,2}
\and
G.Vettolani\inst{2}
\and
A.Cappi\inst{1}
\and
R.Merighi\inst{1}
\and
M.Mignoli\inst{1}
\and
G.M.Stirpe\inst{1}
\and
H.MacGillivray\inst{3}
\and
C.Collins\inst{4}
\and
C.Balkowski\inst{5}
\and
V.Cayatte\inst{5}
\and
S.Maurogordato\inst{5}
\and
D.Proust\inst{5}
\and
G.Chincarini\inst{6,7}
\and
L.Guzzo\inst{6}
\and
D.Maccagni\inst{8}
\and
R.Scaramella\inst{9}
\and
A.Blanchard\inst{10}
\and
M.Ramella\inst{11}
}
%
%
\institute{ 
Osservatorio Astronomico di Bologna, 
via Zamboni 33, 40126 Bologna, Italy
\and
Istituto di Radioastronomia del CNR, 
via Gobetti 101, 40129 Bologna, Italy
\and
Royal Observatory Edinburgh, 
Blackford Hill, Edinburgh EH9 3HJ, United Kingdom
\and
School of EEEP, Liverpool John--Moores University, 
Byrom Street, Liverpool L3 3AF, United Kingdom
\and
Observatoire de Paris, DAEC, Unit\'e associ\'ee au CNRS, D0173 et \`a
l'Universit\'e Paris 7, 5 Place J.Janssen, 92195 Meudon, France
\and
Osservatorio Astronomico di Brera, 
via Bianchi 46, 22055 Merate (LC), Italy
\and
Universit\`a degli Studi di Milano, 
via Celoria 16, 20133 Milano, Italy
\and
Istituto di Fisica Cosmica e Tecnologie Relative, 
via Bassini 15, 20133 Milano, Italy
\and
Osservatorio Astronomico di Roma, 
via Osservatorio 2, 00040 Monteporzio Catone (RM), Italy
\and
Universit\'e L. Pasteur, Observatoire Astronomique, 
11 rue de l'Universit\'e, 67000 Strasbourg, France
\and
Osservatorio Astronomico di Trieste, 
via Tiepolo 11, 34131 Trieste, Italy
}
%
%
\offprints{Elena Zucca (zucca@astbo1.bo.cnr.it)}
\date{Received 00 - 00 - 0000; accepted 00 - 00 - 0000}
\maketitle
\markboth {E.Zucca et al.: 
The ESP galaxy redshift survey: II. The luminosity function and mean galaxy
density}{}
%
%
\begin{abstract}
The ESO Slice Project (ESP) is a galaxy redshift survey we have recently
completed as an ESO Key--Project over about 23 square degrees, in a region 
near the South Galactic Pole. The survey is nearly complete to the limiting 
magnitude $b_J=19.4$ and consists of 3342 galaxies with 
reliable redshift determination.
\\
The ESP survey is intermediate between shallow, wide angle samples and very
deep, one--dimensional pencil beams: spanning a volume of $\sim 5 \times 
10^4$ \htre at 
the sensitivity peak ($z \sim 0.1$), it provides an accurate determination of
the ``local" luminosity function and the mean galaxy density. 
\\
We find that, although a Schechter function (with $\alpha=-1.22$,
$M^*_{b_J}=-19.61$ \hmagp and $\phi^*=0.020$ \hmentre) is an acceptable 
representation of the luminosity function over the entire range of magnitudes 
($M_{b_J}\le-12.4$ \hmagp), our data suggest the presence of a steepening of 
the luminosity function for $M_{b_J}\ge -17$ \hmagp. 
Such a steepening at the faint end of the luminosity function, well fitted 
by a power law with slope $\beta \sim -1.6$, is almost 
completely due to galaxies with emission lines: in fact, dividing our galaxies 
into two samples, i.e. galaxies with and without emission lines, we find 
significant differences in their luminosity functions. In particular, galaxies 
with emission lines show a significantly steeper slope and a fainter $M^*$.
\\
The amplitude and the $\alpha$ and $M^*$ parameters of our luminosity
function are in good agreement with those of the AUTOFIB redshift survey
(Ellis et al. 1996). Viceversa, our amplitude is significantly higher,
by a factor $\sim 1.6$ at $M \sim M^*$, 
than that found for both the Stromlo-APM (Loveday et al. 
1992) and the Las Campanas (Lin et al. 1996) redshift surveys. 
Also the faint end slope of our luminosity 
function is significantly steeper than that found in these two surveys.
\\
The galaxy number density for $M_{b_J}\le -16$ \hmagp is well determined 
($\bar n = 0.08 \pm 0.015$ \hmentre). 
Its estimate for $M_{b_J}\le -12.4$ \hmagp is more 
uncertain, ranging from  $\bar n = 0.28$ \hmentre, in the case of a fit with a 
single Schechter function, to $\bar n = 0.54$ \hmentre, in the case of 
a fit with a Schechter function and a power law.
The corresponding blue luminosity densities in these three cases are 
$\rho_{LUM}= (2.0, 2.2, 2.3) \times 10^8\ h$ \lsole Mpc$^{-3}$, respectively.
\\
Large over-- and under-- densities are clearly seen in our data. In particular,
we find evidence for a ``local" under--density ($n \sim 0.5 \bar n$ for
$D_{comoving}\le 140$ \h) and a significant overdensity ($n \sim 2 \bar n$)
at $z \sim 0.1$. When these radial density variations are taken into account,
our derived luminosity function reproduces very well the observed
counts for $b_J \le 19.4$, including the steeper than Euclidean slope for 
$b_J \le 17.0$. 
\end{abstract}
%
%
\section{Introduction}

An unbiased and detailed characterization of the luminosity function of field 
galaxies is a basic requirement in many extragalactic problems.
Although several determinations of this function are already available in the
literature, the debate about the faint end slope and the normalization of the
luminosity function is still open. These quantities allow a ``local" 
normalization which is crucial for the study of galaxy
evolution and for the explanation of the faint galaxy counts.
\\
Wide angle, shallow samples such as CfA2 ($m_Z \leq
15.5$, Marzke et al. 1994), SSRS2 ($m_{B(0)} \leq 15.5$, da Costa et al. 1994,
Marzke \& da Costa 1997) 
and Stromlo-APM ($b_J \leq 17.15$, Loveday et al. 1992) surveys, 
can provide good determinations of the shape of the luminosity function 
both globally and for different morphological types, but the normalization
can in principle be significantly affected by local fluctuations.
\\
Very deep samples, such as the AUTOFIB composite survey ($b_J \leq 24$,
Ellis et al. 1996) and the CFHT redshift survey ($17.5<I_{AB}<22.5$,
Lilly et al. 1995), are mainly designed to study evolutionary 
effects in the luminosity function.
\\
The ESO Slice Project (ESP, $b_J\leq 19.4$, Vettolani et al. 1997a) is aimed 
to fill the gap between shallow, wide angle surveys and very deep, 
one--dimensional pencil beams, allowing a robust estimate of the faint end 
slope and normalization of the luminosity function derived from a large,
uniform and complete sample of galaxies.
The redshift distribution of the galaxies of this sample, which peaks
at $z \sim 0.1$, is deep enough to allow the sampling of a statistically
representative distribution of the structures in a region of the Universe
where the evolutionary effects are not expected to be important.
\\
The recently published Las Campanas Redshift Survey (LCRS, $r \la 17.5$, 
Lin et al. 1996), based on a large number of galaxies with redshift ($N_{gal} 
\sim 19000$), probes essentially the same redshift depth as the ESP survey, 
but on a larger solid angle. The main differences between the LCRS and
our survey, as well as most of the existing surveys, are the
photometric band in which the galaxies have been selected (red instead
of blue) and the fact that the low surface brightness galaxies have
been deliberately excluded from the spectroscopic sample (Shectman et al.
1996). These facts can make somewhat difficult a direct comparison
between the LCRS luminosity function and others, including ours.
\\
The organization of the paper is as follows. In Sect.2 we briefly summarize
the characteristics of the ESP galaxy redshift survey 
and in Sect.3 we deal with the problem of the
K--correction. In Sect.4 and 5 we describe the methods of derivation and the
results for the luminosity function and the mean galaxy density, respectively.
In Sect.6 we discuss our results and compare them with previous 
estimates of the luminosity function. In Sect.7 we summarize our results.

%
\section{The ESO Slice Project}

The ESO Slice Project (ESP) galaxy redshift survey is described 
in Vettolani et al. (1997a, hereafter paper I) and all the data for the sample 
will be published in Vettolani et al. (1997b, hereafter paper III): here
we summarize only the main characteristics of the survey.
\\
The ESP survey extends over a strip of $\alpha \times \delta = 22^o \times 
1^o$, plus a nearby area of $5^o \times 1^o$, five degrees west of the main 
strip, in the South Galactic Pole region. The position was chosen in order
to minimize the galactic absorption effects ($-60^o \la b^{II} \la -75^o$).
The target objects, with a limiting magnitude $b_J \leq 19.4$, were selected 
from the Edinburgh--Durham Southern Galaxy Catalogue (EDSGC, Heydon--Dumbleton 
et al. 1988).
\\
The right ascension limits are $ 22^{h} 30^m$ and $ 01^{h} 20^m $, at a mean 
declination of $ -40^o 15'$ (1950). We have covered this region with a regular 
grid of adjacent circular fields, with a
diameter of 32 arcmin each, corresponding to the field of view of the multifiber
spectrograph OPTOPUS (Lund 1986) at the 3.6m ESO telescope. The total solid 
angle of the spectroscopic survey is 23.2 square degrees.
In order to increase the overall completeness, in the last observing run
we have used the MEFOS spectrograph (Felenbok et al. 1997), which has a larger
field of view but a smaller number of fibers, to observe some of the
objects which either had not been observed in the OPTOPUS observing runs or
for which only poor quality spectra were available.
\\
We observed a total of 4044 objects, corresponding to $\sim 90\%$ of the 
parent photometric sample, which contains 4487 objects: the objects we
observed were selected to be a random subset of the total catalogue with
respect to both magnitude and surface brightness.
The total number of confirmed galaxies with reliable redshift measurement
is 3342, while 493 objects turned out to be stars and 1 object is a quasar
at redshift $z \sim 1.174$. No redshift measurement could be obtained for the
remaining 208 spectra. 
All the details about data reduction, number of objects and completeness of the 
sample are reported in paper I and paper III. 
\\
The volume of the survey is $\sim 5 \times 10^4$ \htre at $z \sim 0.1$,
corresponding to the sensitivity peak of the survey, and $\sim 1.9\times 10^5$ 
\htre at $z \sim 0.16$, corresponding to the effective depth of the sample.
\\
The absolute magnitudes are computed as
\begin{equation}
M = m - 25 - 5 \log D_L(z) - K(z)
\label{eq:magass}
\end{equation}
where the luminosity distance $D_L$ is given by the Mattig (1958)
expression (throughout the paper we adopt $H_o = 100$ km/s Mpc$^{-1}$ and 
$q_o = 0.5$) and $K(z)$ is the K--correction.
The galactic absorption is assumed to be negligible, because the strip is
close to the South Galactic Pole (Fong et al. 1987). 
\\
Finally, we note that the median internal velocity error for our galaxies 
($\sim 60$ km/s) is quite small compared to the survey depth, and therefore 
its effect on the calculation of absolute magnitudes can be neglected.

%
\section{The K--correction}

Since our database was selected in the blue--green band, K--corrections are 
needed
to compute the luminosity function even for the moderate redshifts sampled
by our galaxies ($z \leq 0.3$).
The application of the functional forms of the K--correction as a function of 
redshift (e.g. Shanks et al. 1984) obviously requires the knowledge of the 
morphological type of each galaxy, which is not available for most of our 
galaxies. We are currently refining a procedure to estimate the K--correction 
for each galaxy directly from the spectra, but this work is still in progress.
Therefore, for this paper we have adopted the following statistical approach.
%
\begin{figure}
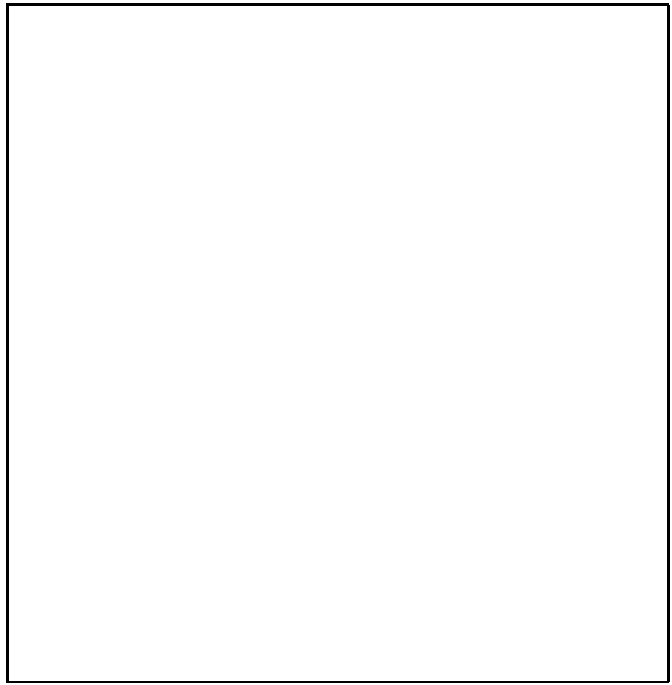

\picplace{9.0cm}
\caption[]{ Adopted weighted K--correction as a function of redshift.
The solid line is a polynomial fit to the points. }
\end{figure}
\\
We have computed an average K--correction, as a function of redshift, 
``weighted" with the expected morphological mix at each redshift. This
average K--correction is defined as
\begin{equation}
< K(z) > = \sum_i f_i (z) K_i (z) 
\label{eq:kz}
\end{equation}
where $f_i (z)$ is the fraction of galaxies of the $i^{th}$ morphological type,
at redshift $z$, whose K--correction is $K_i (z)$.
The expected fractions of each morphological type, as well as their 
K--corrections, have been computed using the Pozzetti et al. (1996) pure
luminosity evolution model. This model, based on the galaxy spectral 
library of Bruzual \& Charlot (1993), is 
constrained to match the empirical K--corrections and the morphological mix
derived locally, and at our moderate redshift the results are
almost independent of the details of the adopted evolutionary model.
In Fig.1 we show our ``weighted" K--correction, computed in bins of 0.025 in
redshift (solid circles), with a polynomial fit superimposed on the points
(solid line).
\\
The decrease of $K(z)$ for $z\ge 0.2$ is essentially due to the fact that
the expected fraction of early--type galaxies, which have a larger 
K--correction, becomes smaller at this redshift, while the fraction of
bluer galaxies with smaller or almost zero K--correction increases. 
\\
Each galaxy at redshift $z'$ is therefore assigned this ``weighted" 
K--correction when computing its absolute magnitude. 
In the non--parametric derivation of the luminosity function (see Sect.4.1) 
we compute for each galaxy the maximum redshift within which it would
still be included in the sample ($z_{max}$). For this purpose the K--correction
for $z' \leq z \leq z_{max}$ is computed by keeping constant the morphological 
mix at the value corresponding to the measured redshift $z'$. 
In this way the K--correction at $z_{max}$ can be written as:
\begin{equation}
< K(z_{max}) > = \sum_i f_i (z') K_i (z_{max}) 
\label{eq:kzmax}
\end{equation}

%
\section{The luminosity function}

%
\subsection{The method}

For most samples of field galaxies the luminosity function is well represented 
by a Schechter (1976) form
\begin{equation}
    \phi(L,x,y,z) dL\ dV = \phi^* \left({L\over{L^*}}\right)^{\alpha}
    {\rm e}^{-L/L^*} d \left({L\over{L^*}}\right)\ dV
\label{eq:sch}
\end{equation}
where $\alpha$ and $L^*$ are parameters referring to the shape of the function
and $\phi^*$ contains the information about the normalization; these parameters
have to be determined from the data.
%
\begin{figure*}
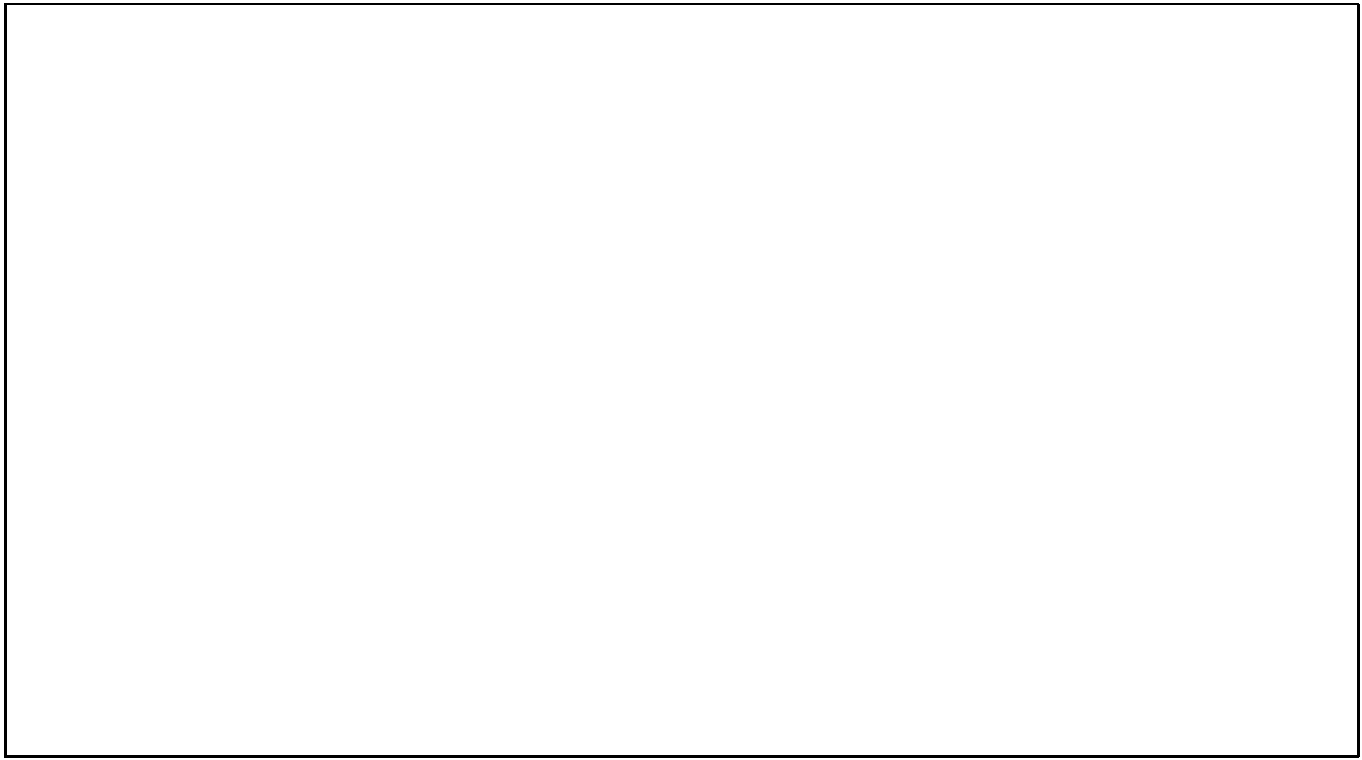

\picplace{10.0cm}
\caption[]{ a) Normalized luminosity function for 3342 ESP galaxies brighter
than $M_{b_J}=-12.4$ \hmagp. The solid circles are computed with a modified 
version of the C--method (error bars represent $1\sigma$ Poissonian 
uncertainties), while the fits are obtained with the STY method. Dashed line: 
single Schechter function; solid line: Schechter function and power law.
b) The same as panel a), but for galaxies with (open squares and dashed 
line) and without (filled squares and dotted line) emission lines. For clarity
only the fit with Schechter function and power law is shown.}
\end{figure*}
\\
Many different methods have been used in the past years to compute
the parameters of the galaxy luminosity function. Recently, however, the STY
method (Sandage et al. 1979) has been the most widely used, and
it has been shown that it is unbiased with respect to density 
inhomogeneities (see for instance Efstathiou et al. 1988, 
Bardelli et al. 1991). The basic idea of this method is to 
compute the estimator of the quantity $\displaystyle{ {{\phi}\over{\Phi}} }$, 
where $\Phi$ is the integral luminosity function. Under the assumption that 
the shape of the luminosity function is not a function of position [i.e. 
$ \phi(L, x, y, z)\ dL\ dV = \rho(x, y, z) dV\ \psi(L) dL $], the probability 
of seeing a galaxy of luminosity $L_i$ at redshift $z_i$ is
\begin{equation}
p_i =  { {\psi(L_i)}\over
{\displaystyle{\int_{L_{min}(z_i)}^{+\infty} \psi(L) dL }}}   
\label{eq:prob}
\end{equation}
where $L_{min}(z_i)$ is the minimum luminosity observable at redshift $z_i$ in
a magnitude--limited sample.
\\
The best parameters $\alpha$ and $L^*$ of the luminosity function
are then determined by maximizing the likelihood function 
${\cal L} (\alpha, L^*)$, which is
the product over all the galaxies of the individual probabilities $p_i$.
This corresponds to minimize the function
\begin{eqnarray}
{\cal S}= -2 \ln {\cal L} = -2 \left[ \alpha \sum_{i=1}^N \ln L_i 
- N (\alpha + 1) \ln L^* -   \right.      \nonumber \\
\left.
{1 \over {L^*}}  \sum_{i=1}^N L_i -   
\sum_{i=1}^N \ln \Gamma\left(\alpha+1, {{L_{min}(z_i)}\over{L^*}} \right)
\right]  
\label{eq:ver}
\end{eqnarray}
where $\Gamma$ is the incomplete Euler gamma function and $N$ is the 
total number of galaxies in the sample. 
\\
The normalization parameter $\phi^*$ is not determined from these equations,
and will be derived from the mean galaxy density in Sect.5.
\\
The STY method is parametric, i.e. it assumes a shape for the luminosity
function: the derived fit can be compared with the results of a non--parametric
method based on the same assumptions.
We use a modified version of the C--method of Lynden--Bell (1971): given a 
galaxy of luminosity $L_i$ at redshift $z_i$ its contribution to the 
luminosity function can be estimated as 
\begin{equation}
\psi(L_i) = {\displaystyle{1-\sum_{k=1}^{i-1} \psi(L_k)}\over
\displaystyle{C^+ (L_i)} }
\label{eq:cmeth}
\end{equation}
where $C^+(L_i)$ is the number of galaxies with luminosity $L \ge L_i$ and
redshift $z \le z_{max}(L_i)$. The resulting integral function 
$\sum_i \psi(L_i)$ is normalized to unity at the minimum luminosity of the
sample. The quantity $\psi(L_i)$ is computed for each galaxy
and the results are binned in luminosity bins and compared with the Schechter
function derived with the STY method. 

\begin{table*}
\caption[]{ Parameters of the luminosity function }
\begin{flushleft}
\begin{tabular}{llllllll}
\hline\noalign{\smallskip}
Sample & $M_{min}$ & $N_{gal}$ & $\alpha$ & $M^*_{b_J}$ & $\phi^*$ (\hmentre) &
$\beta$ & $M_c$ \\
\noalign{\smallskip}
\hline\noalign{\smallskip}
Galaxies in the  & $-12.4$ & $3342$ & $-1.22^{+0.06}_{-0.07}$ & 
                   $-19.61^{+0.06}_{-0.08}$ & $0.020\pm 0.004$ & ~~~~ & ~~~~ \\
total sample     & ~~~~ & ~~~~ & ~~~~ & ~~~~ & ~~~~ & ~~~~ & ~~~~ \\
~~~~             & $-12.4$ & $3342$ & $-1.16$ & $-19.57$ & 
                   $0.021$ & $-1.57$ & $-16.99$ \\
~~~~             & ~~~~ & ~~~~ & ~~~~ & ~~~~ & ~~~~ & ~~~~ & ~~~~ \\
Galaxies with    & $-12.4$ & $1575$ & $-1.40^{+0.09}_{-0.10}$ & 
                   $-19.47^{+0.10}_{-0.11}$ & $0.010\pm 0.002$ & ~~~~ & ~~~~ \\
emission lines   & ~~~~ & ~~~~ & ~~~~ & ~~~~ & ~~~~ & ~~~~ & ~~~~ \\
~~~~             & $-12.4$ & $1575$ & $-1.34$ & $-19.42$ & 
                   $0.010$ & $-1.70$ & $-16.94$ \\
~~~~             & ~~~~ & ~~~~ & ~~~~ & ~~~~ & ~~~~ & ~~~~ & ~~~~ \\
Galaxies without & $-12.4$ & $1767$ & $-0.98^{+0.09}_{-0.09}$ & 
                   $-19.62^{+0.08}_{-0.10}$ & $0.011\pm 0.002$ & ~~~~ & ~~~~ \\
emission lines   & ~~~~ & ~~~~ & ~~~~ & ~~~~ & ~~~~ & ~~~~ & ~~~~ \\
~~~~             & $-12.4$ & $1767$ & $-0.90$ & $-19.57$ & 
                   $0.012$ & $-1.38$ & $-17.22$ \\
~~~~             & ~~~~ & ~~~~ & ~~~~ & ~~~~ & ~~~~ & ~~~~ & ~~~~ \\
\noalign{\smallskip}
\hline
\end{tabular}
\end{flushleft}
\end{table*}

%
\subsection{The results}

The normalized Schechter luminosity function derived with the STY method is 
shown in Fig.2a (dashed line) together with the data points obtained with the 
non--parametric C--method. 
The error bars on these points correspond to the statistical (i.e. Poissonian) 
errors. While the Schechter function is an excellent representation of the 
C--method data points for $M_{b_J} \la -16$ \hmagp, at fainter magnitudes it 
lies below all the points down to $M_{b_J}=-12.4$ \hmagp. We have therefore 
modified the model function, adopting a
Schechter function for $L > L_c$ and a power law for $L \le L_c$.
In this case the low luminosity part of the luminosity function is
described by:
\begin{equation}
    \psi(L) dL = A^* \left({L\over{L^*}}\right)^{\beta} 
           d \left({L\over{L^*}}\right) 
\label{eq:powlaw}
\end{equation}
with $\beta$ and $L_c$ being two additional free parameters in the fitting
procedure, while the normalization $A^*$ is fixed by requiring continuity of 
the two functions at $L=L_c$.
%
\begin{figure*}
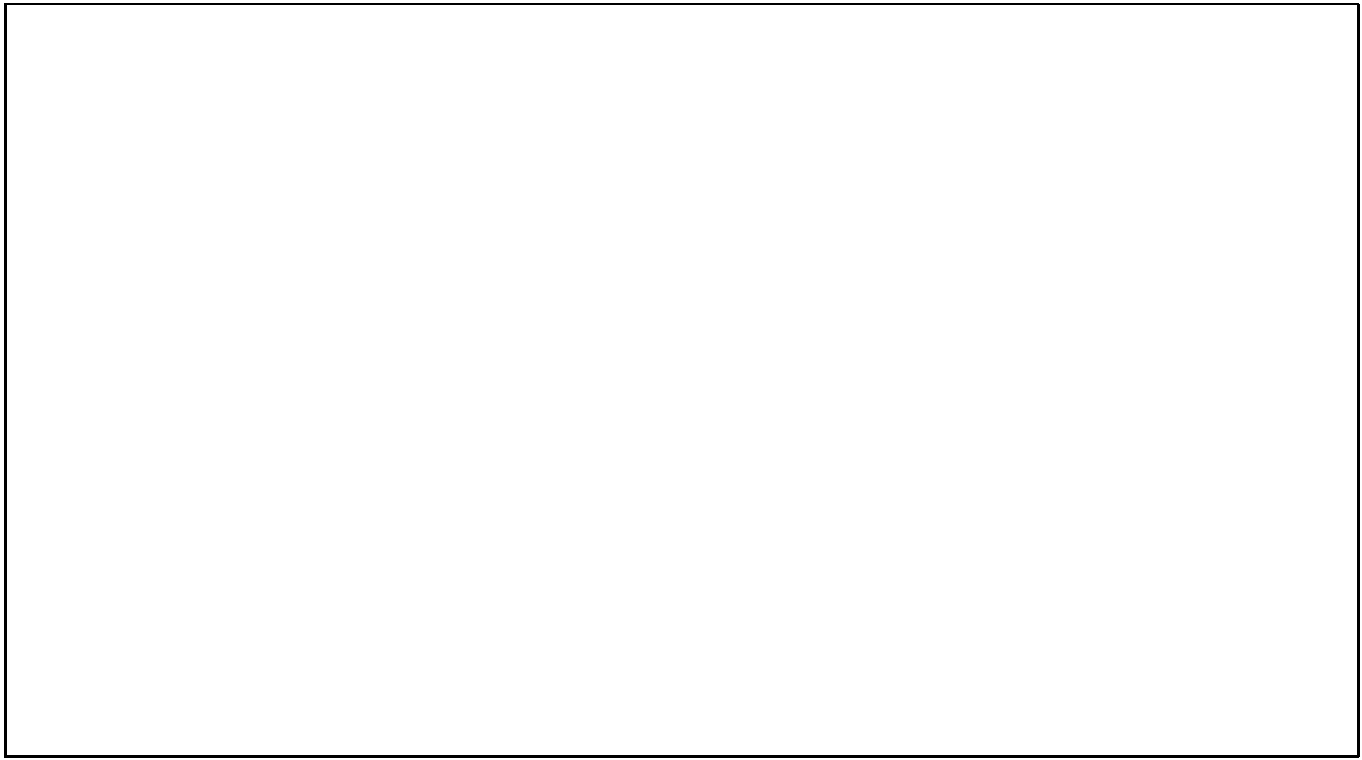

\picplace{10.0cm}
\caption[]{ a) Confidence ellipses at $1\sigma$ and $2\sigma$ levels for the
parameters $\alpha$ and $M^*$, in the case of a fit with a single Schechter 
function, for the total sample and for galaxies with and without emission lines.
b) Confidence ellipse at $2\sigma$ level in the ($\alpha$, 
$\beta$) plane, in the case of a fit with Schechter function and power law for 
the total sample. The dotted line corresponds to the locus $\alpha=\beta$. }
\end{figure*}
\\
The fit with this two--law function (solid line in Fig.2a) is almost
indistinguishable from the single Schechter function at bright magnitudes, and 
reproduces very well also the faint part of the luminosity function. 
Statistically, as judged from the decrease of the ${\cal S}$ function, the 
improvement of the two--law fit with respect to a single Schechter function is 
significant at about $2\sigma$ level, and we can therefore conclude that our 
data suggest the presence of a steepening of the luminosity function 
for $M_{b_J}\ga -17$ \hmagp. In our sample there are 134 galaxies with
$M_{b_J}\ge -17$ \hmagp and 38 galaxies with $M_{b_J}\ge -16$ \hmagp: 
the sampled volumes are $\sim 1.4 \times 10^4$ \htre for galaxies with
$M_{b_J}= -17$ \hmagp and $\sim 3.7 \times 10^3$ \htre for galaxies with
$M_{b_J}= -16$ \hmagp. 
\\
The derived best fit parameters are listed in the first two lines of
Table 1. The errors given for the case of a single Schechter function are
the projections onto the $\alpha$ and $M^*_{b_J}$ axes of the $1\sigma$ 
confidence ellipse (see Fig.3a) for two interesting parameters ($\Delta 
{\cal S}=2.30$).
\\
The errors for the two--law function are not reported in the table because,
given the intrinsic correlations among the four fitted parameters, it is not
very meaningful to give the projection on the four axes of the global 
four--dimensional error volume. However, we analyzed all the possible
combinations of pairs of parameters, considering in turn each pair
as ``interesting'' parameters (Avni 1976). 
From this analysis we find that the errors in $\alpha$ and $M^*$ are
very similar to, although slightly larger than, those computed for the
case of a single Schechter function, while $M_c$ is not well constrained
by the data, with a $1\sigma$ error larger than one magnitude. 
\\
Fig.3b shows the 2 $\sigma$ contours in the ($\alpha$, $\beta$)
plane, which is the most relevant one in determining the reality of the
steepening at low luminosity. Note that no point of the dotted line,
corresponding to the locus $\alpha = \beta$ (i.e. no steepening) is
within the allowed region.
\\
The faint end steepening is almost completely due to galaxies with detectable 
emission lines. In fact, dividing the galaxies into two samples, 
i.e. galaxies with and without emission lines
(1575 and 1767 galaxies respectively), we find very significant
differences in their luminosity functions. Note that, given the typical
signal to noise ratio of our spectra, detection of a line implies
an equivalent width larger than about 5 \AA. Fig.2b shows the normalized
luminosity functions for galaxies with (open squares and dashed line) and
without emission lines (solid squares and dotted line): for clarity only 
the fit for the two--law function (Schechter function and power law at low
luminosity) is plotted. It is clearly seen from the figure that
galaxies with emission lines show a significantly steeper faint end slope
and a slightly fainter $M^*$ (see best fit parameters in Table 1). 
The volume density of galaxies with emission lines is lower than the volume
density of galaxies without emission lines at bright magnitudes, but becomes
higher at faint magnitudes. 
The difference between the luminosity functions of galaxies
with and without emission lines was noted quite early in the project
(see for instance Vettolani et al. 1992), when only a fraction of the
redshifts was available. 
Fig.3a, which shows the $1\sigma$ and $2\sigma$ confidence ellipses of the
parameters $\alpha$ and $M^*$ (single Schechter function case) for the
total sample and for galaxies with and without emission lines, clearly
demonstrates that the difference between the two subsamples is highly
significant. 
%
\begin{figure}
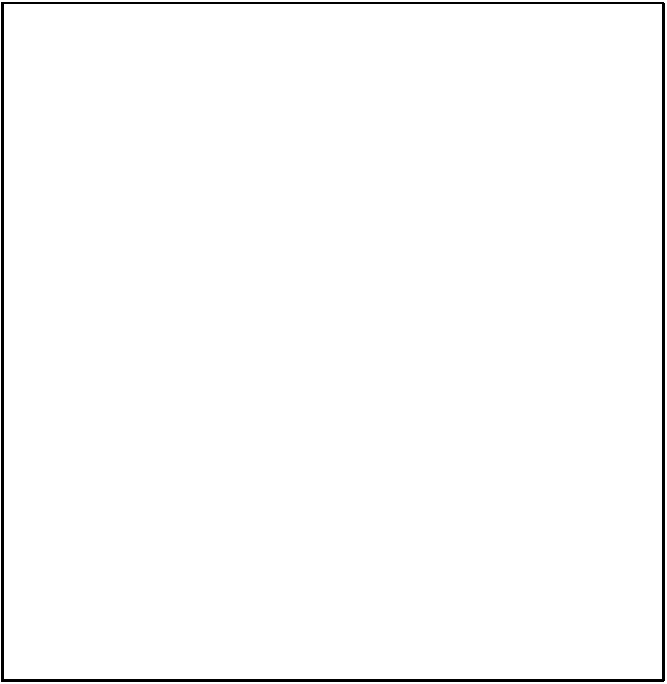

\picplace{9.0cm}
\caption[]{ Confidence ellipses at $1\sigma$ level for the
parameters $\alpha$ and $M^*$ of the luminosity functions in three different
distance bins. The asterisk refers to the best fit parameters for the total 
sample. }
\end{figure}
\\
A similar difference in the best fit parameters of galaxies with and without 
emission lines has been found in the LCRS (see Fig.10 in Lin et al. 1996), 
although for each subsample their best fit slope is significantly 
flatter than the corresponding slope in our survey.
\\
Finally, in order to check the possible existence of evolutionary effects also
at our moderate redshifts, we have divided the ESP galaxies in three comoving 
distance bins, chosen in order to have three samples with a comparable number 
of objects ($D_{com}\le 250$ \h; $250$ \h $< D_{com} \le 360$ \h; $D_{com} > 
360$ \h). The $1\sigma$ confidence ellipses of the ($\alpha$, $M^*$)
parameters for the three derived luminosity functions are reported in Fig.4.
From the overlap of the ellipses we can say that no evidence for evolution
is seen in our data, in agreement with the conclusion of Ellis et al.
(1996) that the bulk of the evolution sets in beyond $z\sim 0.3$.
\\
As a final check, in order to estimate the maximum amount of uncertainty 
induced by our use of a statistical K--correction, we have also computed 
the parameters of the luminosity function with the two extreme assumptions that
all galaxies are either ellipticals or spirals. Even in these cases, the 
derived parameters are not much different from those listed in Table 1.
The maximum variations are $\Delta\alpha \sim -0.06$ (i.e. steeper slope)
and $\Delta M^* \sim -0.22$ (i.e. brighter $M^*$) when we apply to all
galaxies the K--correction appropriate for elliptical galaxies. Moreover,
in both cases the difference between the parameters for galaxies with and
without emission lines remains highly significant. 

%
\subsection{The influence of magnitude errors and redshift incompleteness.}

The observed luminosity function is a convolution of the true luminosity
function with the magnitude error distribution (Efstathiou et al. 1988).
Assuming that the distribution of the magnitude errors is Gaussian, with 
dispersion $\sigma_M$ (independent of $m$),
the observed luminosity function $\phi_{obs}$ is related to the true luminosity 
function $\phi_{true}$ as
\begin{equation}
\phi_{obs}(M) = { 1 \over {\sqrt{2\pi}\sigma_M} }
\int_{-\infty}^{+\infty} \phi_{true}(M') 
              {\rm e}^{ -(M' - M)^2 / 2 \sigma^2_M } dM' 
\label{eq:magerr}
\end{equation}
Preliminary analysis of CCD data in our survey area
for about 80 galaxies in the magnitude range $16.5 \leq b_J \leq 19.4$ shows a 
linear relation between $b_J(EDSGC)$ and $m_B(CCD)$ over the entire range of 
magnitude, with a dispersion ($\sigma_M$) of about 0.2 magnitudes around the 
fit (Garilli et al. in preparation). 
Since the CCD pointings cover the entire right ascension range of our 
survey, this $\sigma_M$ includes both statistical errors within single
plates and possible plate--to--plate zero point variations. 
With the conservative
assumption that the observed $\sigma_M = 0.2$ is entirely due to errors on 
EDSGC magnitudes, we have computed the parameters for the case of a Schechter 
function using the convolved luminosity function of eq.(\ref{eq:magerr}) in the
likelihood function. The resulting best fit parameters are very similar to those
shown in Table 1, with the slope $\alpha$ becoming flatter by $\sim 0.05$ and 
$M^*$ fainter by $\sim 0.10$.
\\
Even more negligible are the effects on the best fit parameters of our small
redshift incompleteness. Eq.(\ref{eq:prob}) is correct only for a complete, 
unbiased sample in which all galaxies with $m < m_{lim}$ are members of the 
sample or all galaxies with $m < m_{lim}$ have the same probability of being 
members of the sample (as, for example, in a redshift survey with $1/n$
sampling). If this is not the case, it has been shown by Zucca et al. (1994) 
that eqs.(\ref{eq:prob}) and (\ref{eq:ver}) have to be modified as:
\begin{equation}
p_i =  \left( { {\psi(L_i)}\over
{\displaystyle{\int_{L_{min}(z_i)}^{+\infty} \psi(L) dL }}}   
\right)^{w_i}
\label{eq:probcor}
\end{equation}
and 
\begin{eqnarray}
{\cal S}=  -2 \left[ \alpha \sum_{i=1}^N w_i \ln L_i 
    - (\alpha + 1) \ln L^* \sum_{i=1}^N w_i -    \right.      \nonumber \\
\left.
{1 \over {L^*}}  \sum_{i=1}^N w_i L_i -   
\sum_{i=1}^N w_i \ln \Gamma\left(\alpha+1, {{L_{min}(z_i)}\over{L^*}} \right)
\right]  
\label{eq:vercor}
\end{eqnarray}
where $w_i$ is the inverse of the probability that the i$^{th}$ galaxy has of 
being included in the sample.  
\\
For the ESP sample there are two kinds of redshift incompleteness: objects 
which have not been observed and objects whose spectra were not useful for 
redshift determination. We have verified that the former incompleteness 
($\sim 10\%$) is consistent with being random in magnitude. As such, it does 
not affect the derived parameters $\alpha$ and $M^*$, but has to be taken into 
account when determining the luminosity function normalization (see next 
section). On the other hand, the latter incompleteness ($\sim 5\%$) is
higher for fainter objects and is well described by:
\begin{equation}
 f(m)= \left\{ \begin{array}{ll}
                 0              & \mbox{if $m\le 16.2$} \\ 
                 0.023 (m - 16.2)    & \mbox{if $m> 16.2$} 
                 \end{array}
         \right.    
\label{eq:inc}
\end{equation}
Note that the maximum fraction of galaxies for which the spectra did not provide
a useful $z$ determination is $\sim 7\%$ for the faintest galaxies of our
survey ($b_J = 19.4$), for which the corresponding weight 
$w_i = 1 / [1 - f(m_i)]$ is 1.08.
\\
Applying this correction to ESP galaxies, we find that the effect on the best 
fit parameters are completely negligible, i.e. $\Delta\alpha \sim \Delta\beta 
\sim -0.02$, $\Delta M^* \sim \Delta M_c \sim 0.01$.

%
\section{The mean galaxy density} 
%
\subsection{The method}

Given a magnitude limited sample, an unbiased estimator for the mean number
density of galaxies $\bar n$ is
\begin{equation}
\bar n = { {\displaystyle{ \sum_i n_i W(z_i) }} \over
            {\displaystyle{ \int F(z) W(z) {{dV}\over{dz}} dz }} }
\label{eq:dh}
\end{equation}
where $W(z)$ is a weighting function and $F(z)$ is the selection function
of the sample, defined as
\begin{equation}
F(z) ={ {\displaystyle{ \int _{\max [ L_1, L_{min}(z) ] } ^ {+\infty}
        \phi(L) dL } }  \over
   {\displaystyle{ \int _{L_1} ^{+\infty}   
        \phi(L) dL } }  }
\label{eq:sel}
\end{equation}
where $L_1$ is the minimum luminosity considered in the estimate of the 
luminosity function. The selection function represents the ratio between 
the number of galaxies detectable
at redshift $z$ and the total number of galaxies with $L\ge L_1$ at the same
redshift. Davis \& Huchra (1982; DH82) have discussed a number of different
estimators for $\bar n$, corresponding to different choices for the 
weighting function
$W(z)$, and have shown that the minimum variance estimator is $n_{J_3}$,
obtained with $W(z_i) =  1 / [1 + {\bar n} J_3 F(z_i)]$, where $J_3$ is
the second moment of the spatial correlation function $\xi(r)$.
Adopting $J_3\sim 10,000$, as found from a fit of the correlation function in 
various surveys (see for instance Stromlo-APM and LCRS), including ours, and 
using the formula given in the appendix of DH82, we find that the expected 
fractional error on $\bar n$ is $\sim 10\%$. Note, however, that this error 
estimate accounts only for the part due to galaxy clustering and does not take
into account the additional uncertainty arising from the errors on the
parameters of the luminosity function. By varying $\alpha$ and $M^*$ along
their 1 $\sigma$ confidence ellipse, we find that the errors on $\bar n$ 
induced by the uncertainty on the parameters of the luminosity function
are significantly larger ($\sim 20\%$) than those due to galaxy clustering. 
Therefore, since there is no a priori reason to prefer the $n_{J_3}$ estimator
with respect to other possible estimators, we have analyzed the behaviour
of all the estimators discussed by DH82. 
%
\begin{figure}
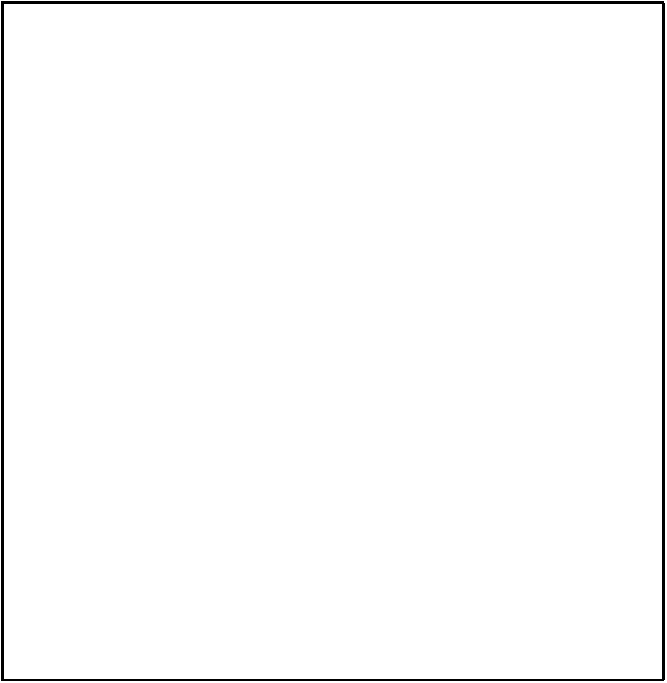

\picplace{9.0cm}
\caption[]{ Integrated average density as a function of the maximum comoving
distance for the four DH82 estimators (see text). }
\end{figure}
%
\begin{figure}
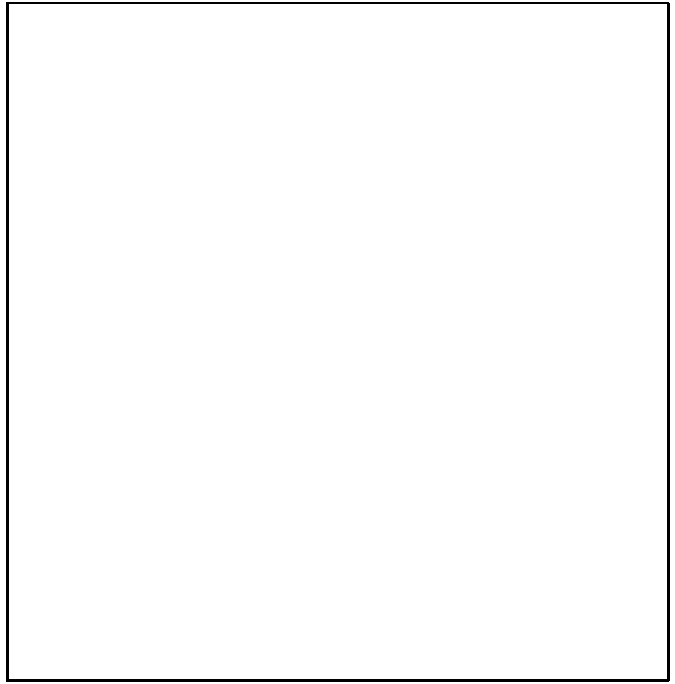

\picplace{9.0cm}
\caption[]{ a) Galaxy density as a function of the comoving distance: distance
shells have been chosen in order to have a number of expected object $\ge 150$;
error bars represent $1\sigma$ Poissonian uncertainties. The solid straight
line corresponds to the mean value $\bar n$, dashed lines represent $1\sigma$
uncertainties. The values refer to galaxies with $M_{b_J}\le-12.4$ \hmagp, 
in the case of a fit with a Schechter function. 
b) Comoving distance histogram of ESP galaxies: the solid line is the
distribution expected for a uniform sample. }
\end{figure}
\\
Fig.5 shows the estimated average density as a function of the maximum 
distance for the four DH82 estimators: the values refer to galaxies with 
$M_{b_J}\le-12.4$ \hmagp, in the case of a fit with a Schechter function.
The relative behaviour of the curves does not change with other choices 
for $M_{min}$ or for the case of the fit with a Schechter function and
a power law.
The $n_3$ estimator produces a very stable value for the galaxy
density for all comoving distances $\ga 400$ \h, even for
distances larger than the maximum distance in the sample.
The difference between the maximum and minimum values of the $n_3$ estimator
for $D_{com} > 400$ \h is of the order of $6\%$, well below the $20\%$
uncertainty due to the errors on $\alpha$ and $M^*$.
All the other estimators appear to be much more sensitive to local
fluctuations and vary significantly with the assumed maximum distance.
For example, the density resulting from the $n_{J_3}$ estimator would change
by $\sim 25\%$ by changing the maximum distance from $\sim 650$ \h, 
within which $\sim 99\%$ of our galaxies are contained, to $\sim 780$ \h,
which corresponds to the maximum distance in the sample. Since an optimal
choice for the maximum distance to adopt in this analysis is not defined, 
this result shows an intrinsic
ambiguity, at least for our sample, connected to the $n_{J_3}$ estimator.
On the basis of this figure we therefore decided to adopt the $n_3$ estimator, 
which corresponds to a constant weighting function in eq.(\ref{eq:dh}).
The same estimator has been used in recent analyses of other redshift
surveys, such as the LCRS (Lin et al. 1996) and the SSRS2 (Marzke \& da Costa
1997).
The errors on ${\bar n}$ are computed applying the error propagation on
eq.(\ref{eq:dh}), taking into account also the uncertainties arising by 
varying $\alpha$ and $M^*$ along their confidence ellipse.
\\
The derived densities need to be corrected for the redshift incompleteness of 
our survey: indeed, among the 4487 objects of the photometric catalogue, we 
have 208 failed spectra with low signal to noise ratio and 443 not observed
objects (see paper I and Sect.4.4). We assume that the failed spectra 
correspond to galaxies, because stars, being point--like objects, have
on average a better signal to noise ratio than galaxies, and 
that the percentage of stars in not observed objects spectra is the same as
in the spectroscopic sample (i.e. $\sim 12.2\%$). After subtracting the 
expected stellar contamination, we are left with about 600 
not observed galaxies, corresponding to $\sim 15\%$ of the total. Therefore
the galaxy counts are multiplied by the factor 1.15, before deriving
the mean density. 
The use of a simple multiplicative factor implicitely assumes that the
distance distribution of non observed galaxies is the same as that of
the whole sample. This assumption is supported by the analysis of
the magnitude distributions discussed in Sect.4.3.
\\
Finally, from the estimated $\bar n$, it is possible to derive the 
normalization of the luminosity function as
\begin{equation}
\phi^* ={ {\bar n} 
\over{\displaystyle { \Gamma\left(\alpha+1,{{L_1}\over{L^*}}\right)       } }
}
\label{eq:phi*1}
\end{equation}
in the case of a single Schechter function, and as
\begin{equation}
\phi^* ={ {\bar n} 
\over{\displaystyle { \Gamma\left(\alpha+1,{{L_c}\over{L^*}}\right) +
\left({{L_c}\over{L^*}}\right)^{\alpha-\beta} 
{ { {\rm e}^{-L_c/L^*}}} 
\int_{L_1}^{L_c} \left({{L}\over{L^*}}\right)^{\beta} 
d \left({L\over{L^*}}\right)
} }
}
\label{eq:phi*2}
\end{equation}
in the case of Schechter function and power law. 

%
\subsection{The results}

The density radial profile of ESP galaxies is shown in Fig.6a
as a function of comoving distance. In each distance bin the
density has been computed using the $n_3$ estimator as discussed above; 
the error bars represent $1\sigma$ Poissonian uncertainties.
The width of the various distance shells (see horizontal lines in the figure)
has been chosen in order to have  $\sim 150$ expected galaxies in each bin.
The solid line represents the value of the global ${\bar n}$ 
derived from the total sample (see Fig.5) and the dashed lines indicate the
$\pm 1\sigma$ uncertainty on this value. The median of the ${\bar n}$ values
in the shells is $0.29 \pm 0.04$ \hmentre, in excellent agreement with the 
global value $\bar n = 0.28\pm 0.05$ \hmentre. 
Although the densities derived from most of the distance shells are within
$\pm 2\sigma$ from the global mean density of the sample, at least three
regions have densities which differ significantly ($\ga$ a factor of two)
from the mean density. These regions (two underdense regions at
$D_{com} \le 140$ \h and $D_{com} \sim 230$ \h, and an overdense region at
$D_{com} \sim 290$ \h) are clearly visible also in Fig.6b, which shows the
observed distance histogram and the distribution expected for a uniform
sample. The relatively large density fluctuations seen in our data are
clearly due to the fact that our survey is a slice with a narrow width in 
one direction. Note, however, that over--densities and under--densities
of about a factor two are seen also in wider angle surveys, such as
the SSRS2 (Marzke \& da Costa 1997), the CfA2 (Marzke et al. 1994)
and the LCRS (Lin et al. 1996) surveys. 
%
\begin{figure*}
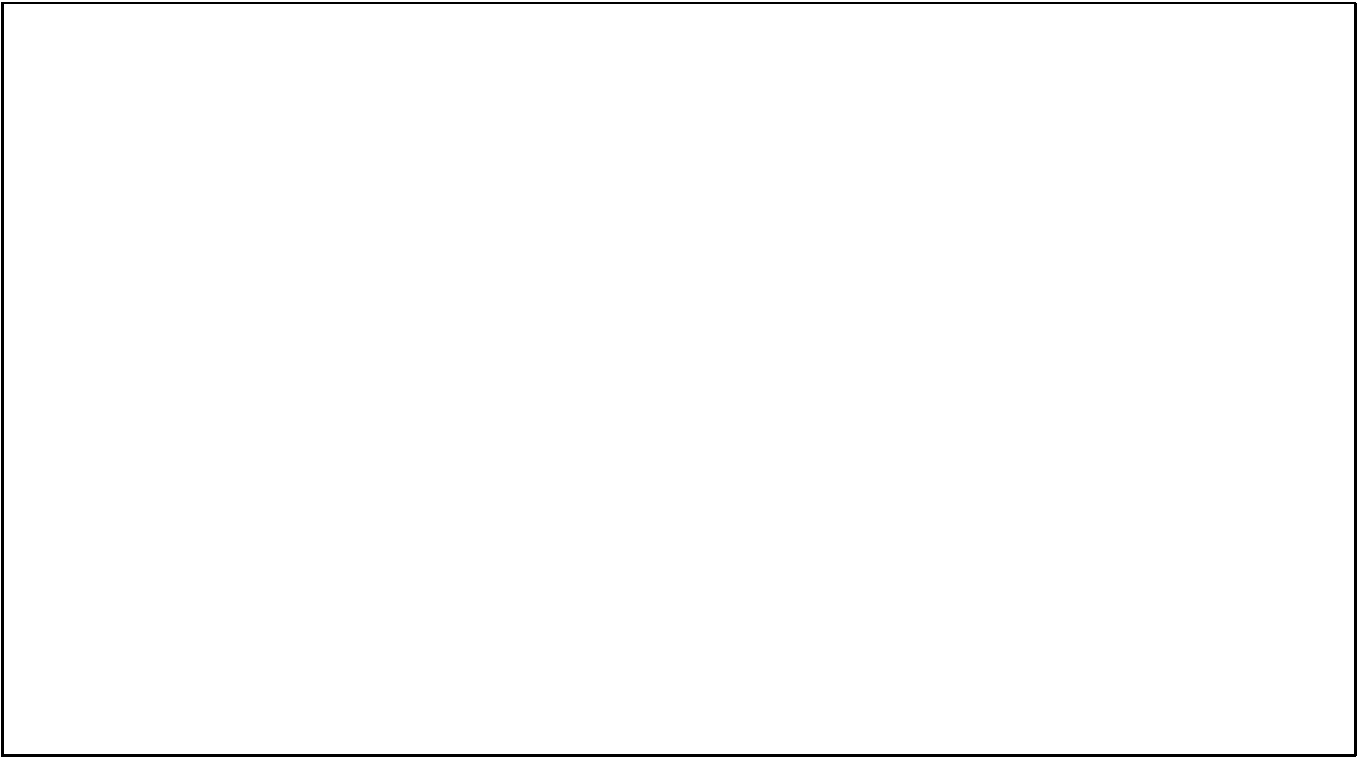

\picplace{10.0cm}
\caption[]{ a) Comparison of the ESP luminosity function with previous results.
b) Ratio between the ESP luminosity function and the others. In this
panel, in order to be consistent with the other luminosity functions, we
have used for the ESP galaxies the fit obtained with a single Schechter
function. The dotted straight lines represent an approximate
$\pm 2\sigma$ uncertainty on the ratio derived from the uncertainties on the
normalizations of the various surveys. For sake of clarity, the two shallow 
surveys (CfA2 and SSRS2) are not plotted. }
\end{figure*}
\\
The results shown in this Fig.6a
refer to the case of a fit with a single Schechter function for galaxies
brighter than $M_{b_J} = -12.4$ \hmagp. Qualitatively, the results for
different limiting absolute magnitudes or for the case of the fit with
a Schechter function and a power law are similar to those shown in Fig.6a, 
except for different normalizations.
The galaxy number density for $M_{b_J}\le -16$ \hmagp is well determined 
($\bar n = 0.08\pm 0.015$ \hmentre) and is essentially independent from
the adopted fitting law for the luminosity function.
Its estimate for $M_{b_J}\le -12.4$ \hmagp is more 
uncertain, ranging from $\bar n = 0.28\pm 0.05$ \hmentre, in the case of a fit 
with a single Schechter function, to $\bar n = 0.54\pm 0.10$ \hmentre, in the 
case of a fit with a Schechter function and a power law.  
The corresponding luminosity densities in these three cases are $\rho_{LUM}= 
(2.0, 2.2, 2.3) \times 10^8\ h$ \lsole Mpc$^{-3}$, respectively: the
similarity of these values indicates that the galaxies in the faint end
of the luminosity function contribute strongly to the {\it number} density, 
but change only slightly the global {\it luminosity} density.
\\
From these number densities we have then derived $\phi^*$ for the various 
cases: the obtained values are reported in Table 1, in the case of 
$M_{min} = -12.4$ \hmagp; the values do not change with other choices for
$M_{min}$. 

\begin{table*}
\caption[]{ Parameters of the luminosity function from various samples }
\begin{flushleft}
\begin{tabular}{llllllll}
\hline\noalign{\smallskip}
Sample & $N_{gal}$ & $m_{lim}$ & $M_{min}$ & $\alpha$ & $M^*$ & 
$\phi^*$ (\hmentre) & notes \\
\noalign{\smallskip}
\hline\noalign{\smallskip}
ESP      & $3342$ & $b_J=19.4$ & $-12.4$ & $-1.22^{+0.06}_{-0.07}$ & 
           $-19.61^{+0.06}_{-0.08}$ & $0.020\pm 0.004$ & ~~~~ \\
CfA2     & $9063$ & $m_Z=15.5$ & $-16.5$ & $-1.0\pm 0.2$ &
           $-18.8\pm 0.3$ & $0.04\pm 0.01$ & ~~~~ \\
SSRS2    & $3288$ & $m_{B(0)}=15.5$ & $-14$ & $-1.16^{+0.08}_{-0.06}$ &
           $-19.45\pm 0.08$ & $0.0109\pm 0.0030$ & ~~~~ \\
Stromlo-APM & $1658$ & $b_J =17.15$ & $-15$ & $-0.97\pm 0.15$ &
            $-19.50\pm 0.13$ & $0.0140\pm 0.0017$ & ~~~~ \\
LCRS     & $18678$ & $r\sim 17.5$ & $-17.5$ & $-0.70\pm 0.05$ & 
           $-20.29\pm 0.02$ & $0.019\pm 0.001$ & $<b_J - r>_o = 1.1$ \\
AUTOFIBa & $588$ & $b_J=24$ & $-14$ & $-1.16^{+0.05}_{-0.05}$ &
           $-19.30^{+0.15}_{-0.12}$ & $0.0245^{+0.0037}_{-0.0031}$ & 
           $0.02<z<0.15$ \\
AUTOFIBb & $665$  & $b_J=24$ & $-16$ & $-1.41^{+0.12}_{-0.07}$  &
           $-19.65^{+0.12}_{-0.10}$ & $0.0148\pm ^{+0.0030}_{-0.0019}$ & 
           $0.15<z<0.35$  \\
\noalign{\smallskip}
\hline
\end{tabular}
\end{flushleft}
\end{table*}

%
\section{Discussion and comparison with previous results}

\begin{figure}
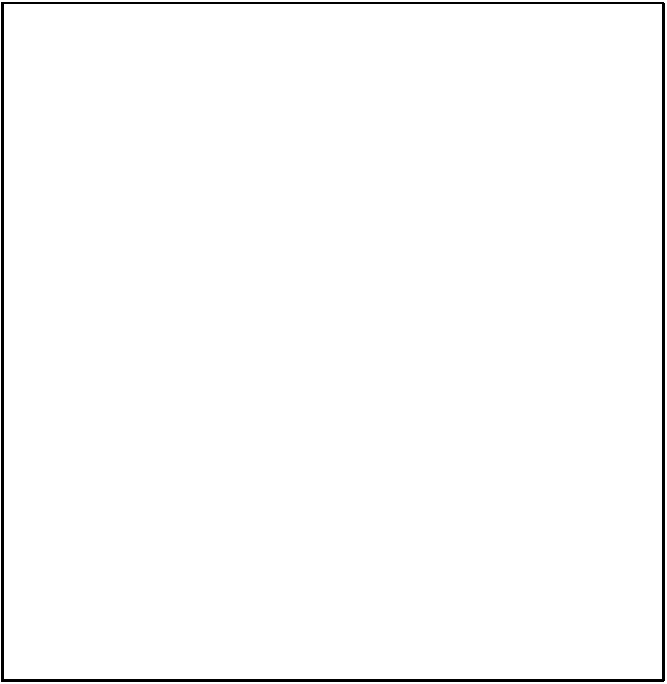

\picplace{9.0cm}
\caption[]{ Observed galaxy counts compared with those expected from the 
luminosity function (solid line); the same but taking into account the observed
radial density variations (dashed line) }
\end{figure}

In Table 2 we list the parameters from the most recent luminosity function
determinations available in the literature.
From this table it is clear that
there are significant differences between the various surveys.
We first note the discrepant value of $M^*$
found in the CfA2 survey (Marzke et al. 1994) with respect to all the other
samples, which cannot be entirely explained by the different passband used. 
The $\alpha$ and $M^*$ parameters derived from the SSRS2 survey (Marzke \&
da Costa 1997) are in good agreement with our results.
However, the amplitude of the SSRS2 luminosity function is such that the ratio 
between the SSRS2 and the ESP luminosity functions is in the range 
$0.4 - 0.5$ for $-20 < M < -14$.
\\
In Fig.7a we compare the luminosity functions derived from the various
surveys; in order to better visualize the differences, we show in Fig.7b
the ratio between the ESP luminosity function and the others. 
\\
From this figure we find that:
\\
i) The AUTOFIB luminosity functions (Ellis et al. 1996) in their first
two redshift bins, which cover approximately the same distance range as
our survey, are in good agreement with the ESP luminosity function. Both
AUTOFIB luminosity functions are within $2\sigma$ from the ESP luminosity
function at all magnitudes.
\\
ii) The Stromlo-APM (Loveday et al. 1992) and the LCRS (Lin et al. 1996) 
luminosity functions are significantly different from the ESP luminosity
function in both shape (both are flatter than the ESP at low luminosity)
and amplitude (both have a lower amplitude by a factor $\sim 1.6$
at $M \sim M^*$). 
\\
While a possible explanation for the difference in 
amplitude with respect to the Stromlo-APM luminosity function is presented 
later in this section, we are intrigued by the significant difference in
amplitude with the LCRS luminosity function, which is derived from
galaxies with a redshift distribution very similar to ours. The fact that
the LCRS galaxies are selected in the red band can in principle explain,
at least in part, their flatter luminosity function on the basis of
the fact that at low luminosity most of the galaxies have emission
lines (see Sect.4.2 above) and are presumably bluer than the average.
The dependence of the local luminosity function on color has been recently 
discussed by Marzke \& da Costa (1997), who indeed found a significantly
steeper slope for blue galaxies with respect to the red ones in the SSRS2.
This, however, should not affect significantly the amplitude at 
$M \sim M^*$. It is true that in the LCRS low surface brightness galaxies have 
been explicitly eliminated from the redshift survey, and therefore from the
computation of the luminosity function, but the fraction of galaxies
eliminated because of this selection effect is estimated to be less
than $10\%$ (Lin et al. 1996). An other difference between the LCRS and the ESP
surveys is in the fraction of stars in the spectroscopic sample of candidate 
galaxies ($\sim 3\%$ in LCRS and $\sim 12\%$ in ESP). This can in principle
be due to a better star--galaxy separation in the Las Campanas photometric 
data. Alternatively, it is possible that a non negligible number of galaxies, 
probably the most compact ones, have been classified as stars in the Las 
Campanas data and therefore have not been observed spectroscopically. 
Obviously, we have no way to check if something like this has 
really happened in the Las Campanas data.
\\
We have used our derived luminosity function to compute the expected
galaxy counts in the case of a constant density and compared this prediction 
(solid line in Fig.8) with the data in our photometric catalogue (solid circles
with error bars in the same figure). 
Note that in this magnitude range ($b_J \le 20$) the counts predicted using
the two different fits (single Schechter function or Schechter function and 
power law) are almost identical, while they differ significantly at magnitudes
fainter than $b_J \sim 25$.
\\
It is clear from the figure that, while the counts predicted from the 
luminosity function are reasonably 
consistent with the data for $b_J \ga 17$, they are significantly higher at 
brighter magnitude. Since our counts, although with large statistical errors 
because of the small area, are consistent with the global EDSGC  
(Heydon--Dumbleton et al. 1989) counts and the APM counts are consistent
with the EDSGC ones (Maddox et al. 1990), this effect is 
essentially the same which has led Maddox et al. (1990) to suggest rapid and 
dramatic evolution in the galaxy properties for $z \la 0.2$. Such a strong 
``local'' evolution is not seen neither in our analysis of the luminosity 
function in three distance intervals (see Sect.4.2) nor in the deeper AUTOFIB
redshift survey from which very little, if any, evolution is seen up to 
$z = 0.35$.
\\
It has been suggested that the low APM counts at bright magnitude may be due 
to a magnitude scale error in the Stromlo-APM galaxy survey (Metcalfe
et al. 1995). The possible existence of such an error for the APM and
similar catalogues is reinforced by the recent analysis of well calibrated
Schmidt plates digitized with the MAMA machine by Bertin and Dennefeld (1997).
We can not exclude a similar problem in our data, although the admittedly
limited CCD photometry we have obtained on bright galaxies in this area
(Garilli et al. in preparation) does not show any strong magnitude error for 
galaxies in the magnitude range $16 - 17$.
\\
At least for our data, however, an alternative or additional interpretation
is suggested by Fig.6, which shows significant fluctuations around the mean 
density. 
The dashed line in Fig.8 shows the expected counts computed by relaxing
the assumption of constant density and allowing the normalization
of the luminosity function to change with distance 
as indicated by the data points in Fig.6. In
this way the agreement between the predicted and observed counts is more
than acceptable over the entire magnitude range. In particular, the deficiency 
of the observed counts at bright magnitudes would be due to the ``local'' 
($D_{com} \le 140$ \h) under--density. 
In this respect, it is interesting to note that, as mentioned
above, the amplitude of the ESP luminosity function is about a factor of two 
higher than the SSRS2 one, which has been derived from galaxies with 
$D_{com} \la 140$ \h over a much larger area, which includes the region of our 
survey. Moreover, an under--density over a similar distance range is seen also 
in the South Galactic Cap part of the LCRS (see Fig.8a and 8c in Lin et
al. 1996). Since the LCRS is based on galaxies extracted from a much
larger area ($80 \times 6.5$ sq.deg.), over which our survey region is fully
contained, we are led to conclude that the local under--density seen in our data
may have a size of more than $100$ \h in at least the right ascension
direction. If such an under--density extends with a similar size also
in the declination direction, it could contribute significantly, possibly 
in addition to magnitude scale errors, to the low APM counts of bright
galaxies. 
\\
In this framework, the difference in amplitude 
between our luminosity function and the Loveday et al. (1992) luminosity
function based on APM galaxies with $b_J \le 17.15$ could be at least partly
explained by the presence of a real, giant under--density extending over a
significant fraction of their survey. If we compute a luminosity function with 
our data but with the same limiting magnitude as in Loveday et al., the 
parameters we derive are fully consistent in both shape and amplitude with 
those derived by them. It has to be stressed, however, that
the possibility for such an explanation has been considered by Loveday et al.,
but they did not find any evidence for it in their data (see, for example,
their Fig.4). 
\\
It is also interesting to note that, since a strong ``local" evolution is
not easily accomodated in the standard evolutionary galaxy models, most
models for the faint galaxy counts have often assumed a high normalization
of the local luminosity function (see for example Guiderdoni \& 
Rocca--Volmerange 1990, Pozzetti et al. 1996). We find that the normalization
of our luminosity function is in good agreement with that usually assumed
in these models and therefore our result gives an {\it a posteriori} 
confirmation of the somewhat {\it ad hoc} assumption adopted in these models.

%
\section{Conclusions}

In this paper we have derived the luminosity function and the mean density
from the ESP galaxy redshift survey; our main results are the following:

1) Although a Schechter function is an acceptable representation of the
luminosity function over the entire range of magnitudes ($M_{b_J}\le-12.4$
\hmagp), our data suggest the presence of a steepening of the luminosity 
function for $M_{b_J}\ge -17$ \hmagp. Such a steepening, well fitted by a 
power law with slope $\beta \sim -1.6$, is in agreement with what has been 
recently found by similar analyses for both field galaxies (Marzke et al. 1994) 
and galaxies in clusters (see for instance Driver \& Phillipps 1996).

2) The steepening at the faint end of the luminosity function is almost 
completely due to galaxies with emission lines: in fact, dividing our galaxies 
into two samples, i.e. galaxies with and without emission lines, we find 
significant differences in their luminosity functions. In particular, galaxies 
with emission lines show a significantly steeper slope and a fainter $M^*$.
The volume density of galaxies with emission lines is lower than the volume
density of galaxies without emission lines at bright magnitudes, but becomes
higher at faint magnitudes. 

3) The amplitude and the $\alpha$ and $M^*$ parameters of our luminosity
function are in good agreement with those of the AUTOFIB redshift survey.
Viceversa, our amplitude is a factor $\sim 1.6$ higher, at $M\sim M^*$,
than that found for 
both the Stromlo-APM and the Las Campanas redshift surveys. Also the faint end 
slope of the luminosity function is significantly steeper for the ESP 
galaxies than that found in these two surveys.

4) We find evidence for a local under--density, extending up to a comoving
distance $\sim 140$ \h. The volume probed by the ESP within such a distance
is smaller than the volume of a typical void with 50 \h diameter, and in
principle this observed nearby under--density could be due to the specific
direction of the survey piercing through a local void. Our data do not
allow to characterize this low density region in terms of size and shape. 
When the radial density variations observed in our
data are taken into account, our derived luminosity function reproduces
very well the observed counts for $b_J \le 19.4$, including the steeper than
Euclidean slope for $b_J \le 17.0$. If this under--density extends over a
much larger solid angle than that covered by our survey, it could, at least
partly, explain the low amplitude of the Stromlo-APM luminosity function.

5) A similar explanation can not justify the significant difference in
amplitude between the ESP and the LCRS luminosity functions, because
the two samples cover essentially the same redshift range. One possibility,
which has however to be verified, is that a non negligible number of
galaxies are missing from the original CCD photometric catalog of the LCRS.

Given the large number of galaxies, the high degree of completeness, 
the accurate selection criteria and the good photometry
of the ESP redshift survey, we can conclude that the
results of our analysis give the best available estimate of 
both the normalization and the faint end slope of
the luminosity function in the local Universe.

%
\begin{acknowledgements}
This work has been partially supported through NATO Grant CRG 920150,  
EEC Contract ERB--CHRX--CT92--0033, CNR Contract 95.01099.CT02 and by 
Institut National des Sciences de l'Univers and Cosmology GDR.
\\ 
We warmly thank Lucia Pozzetti for her help with the weighted K--corrections
discussed in Sect.3. We thank the referee for useful comments.
\end{acknowledgements}
%
%

%
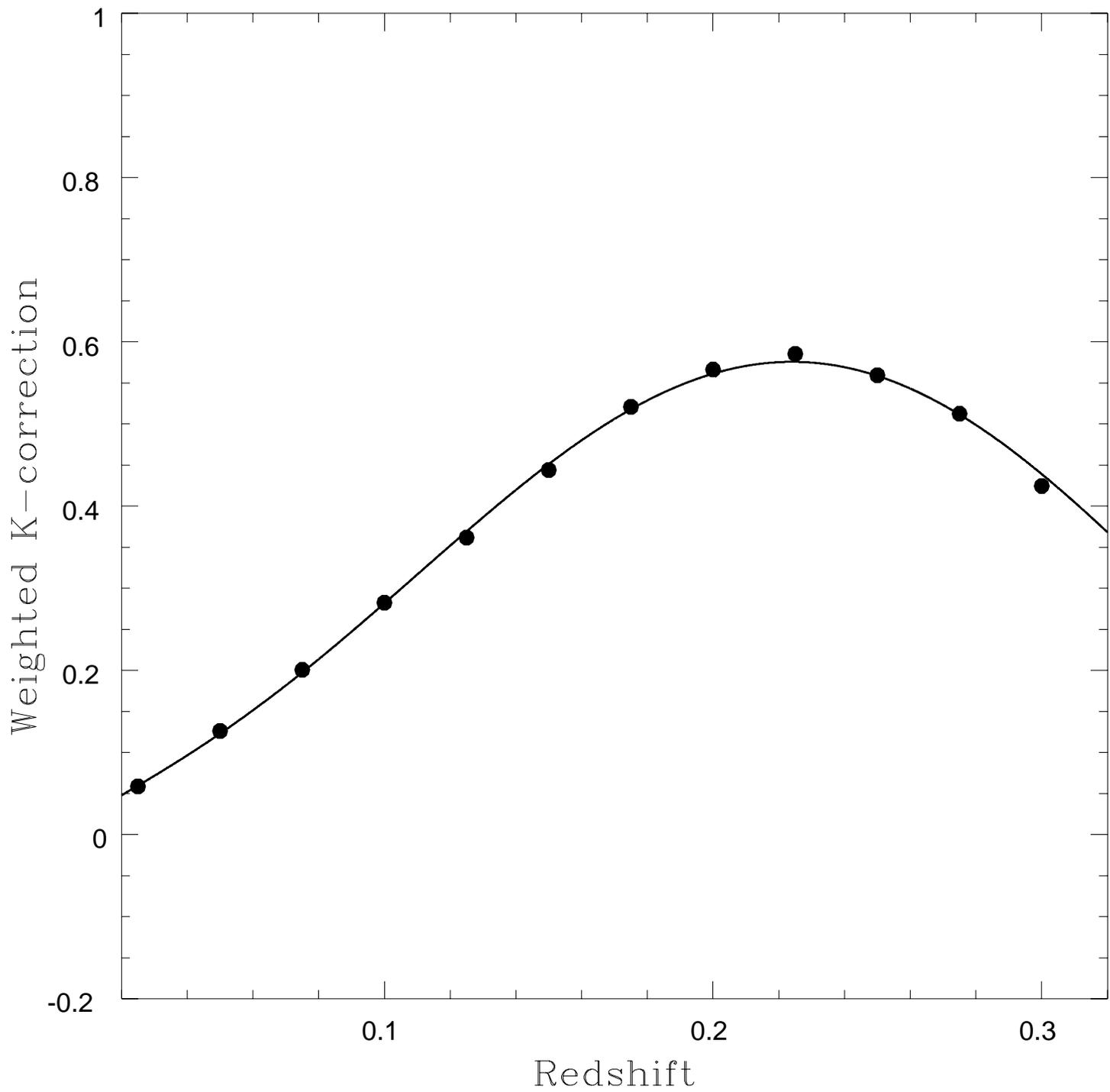
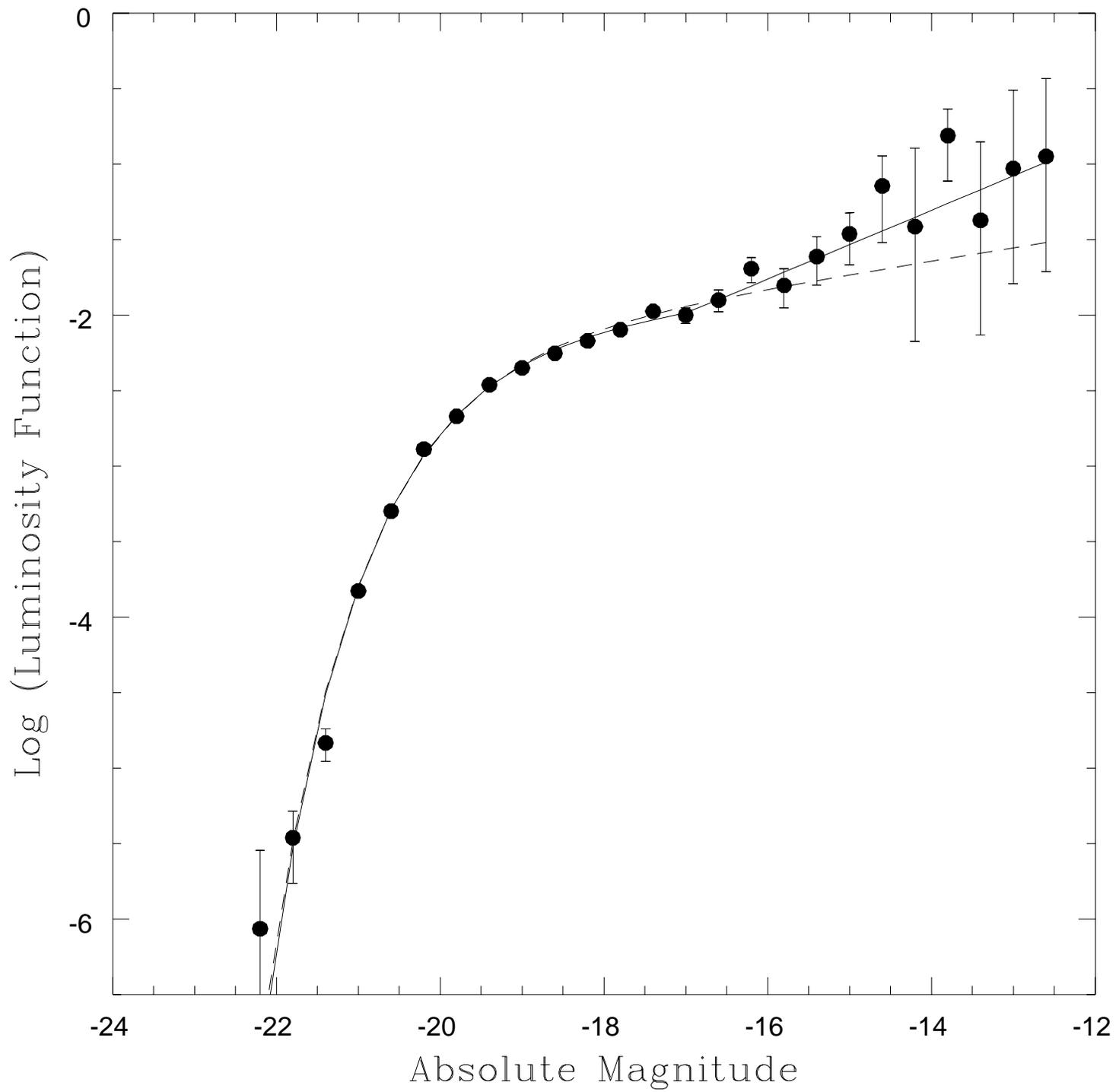
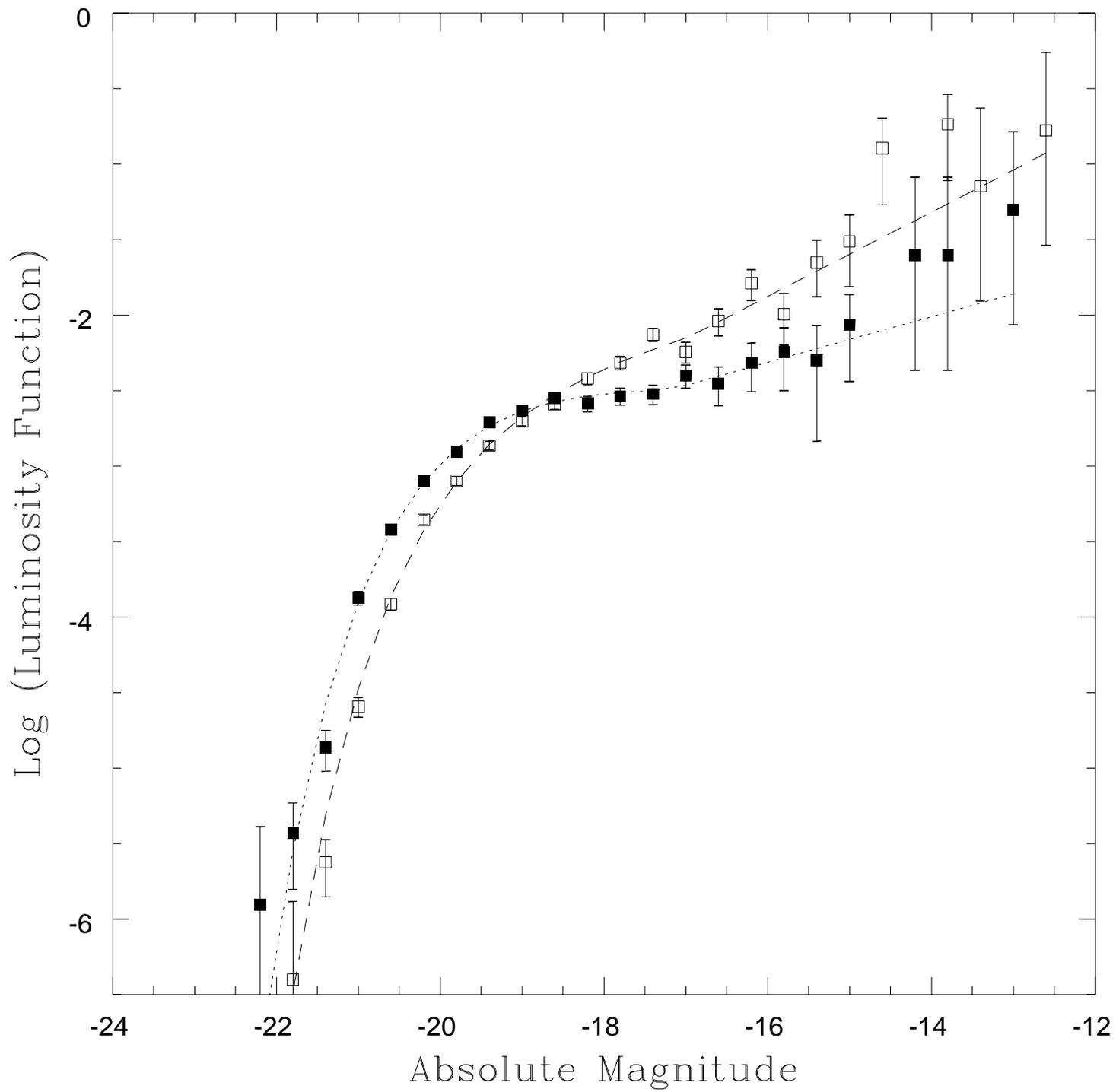
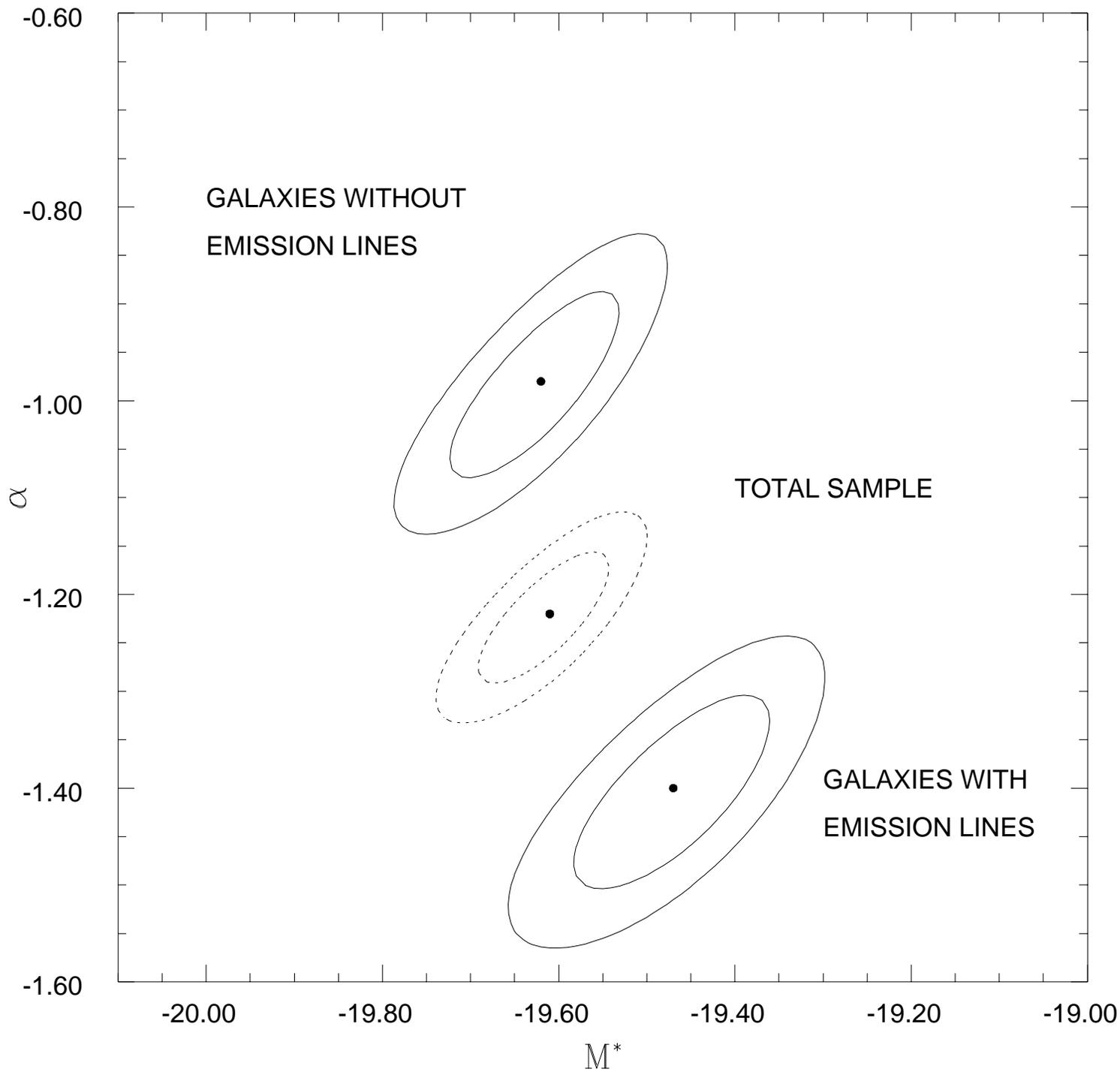
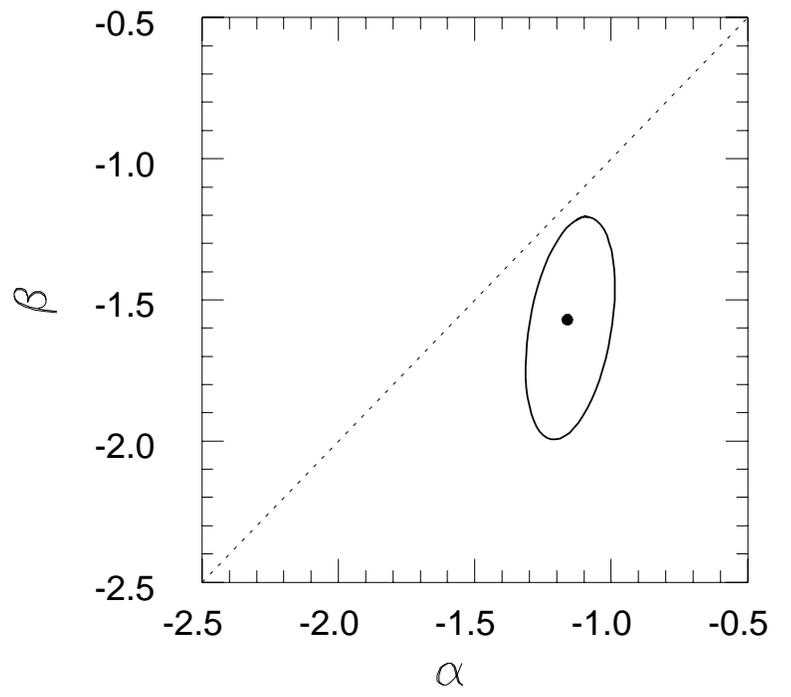
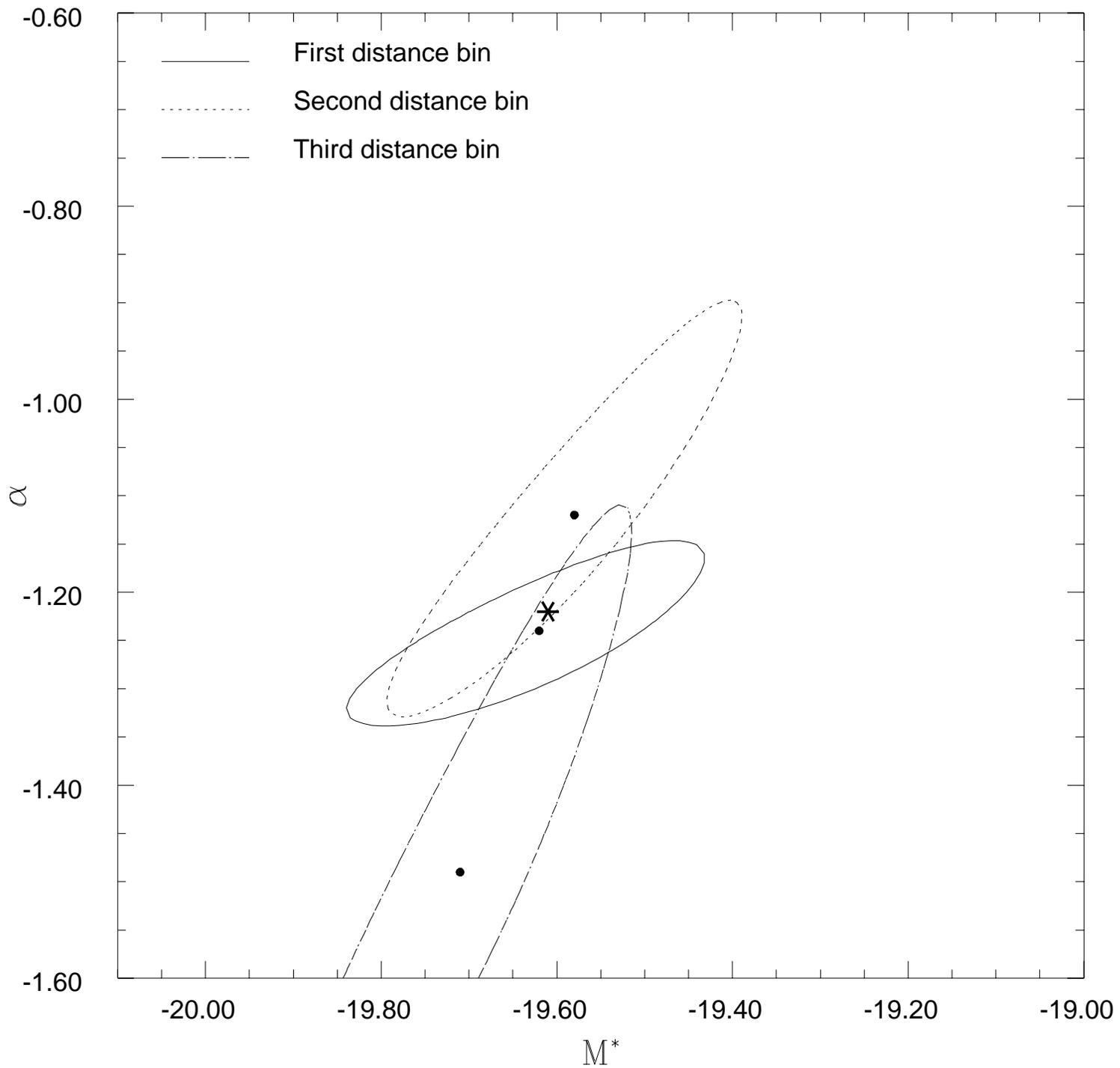
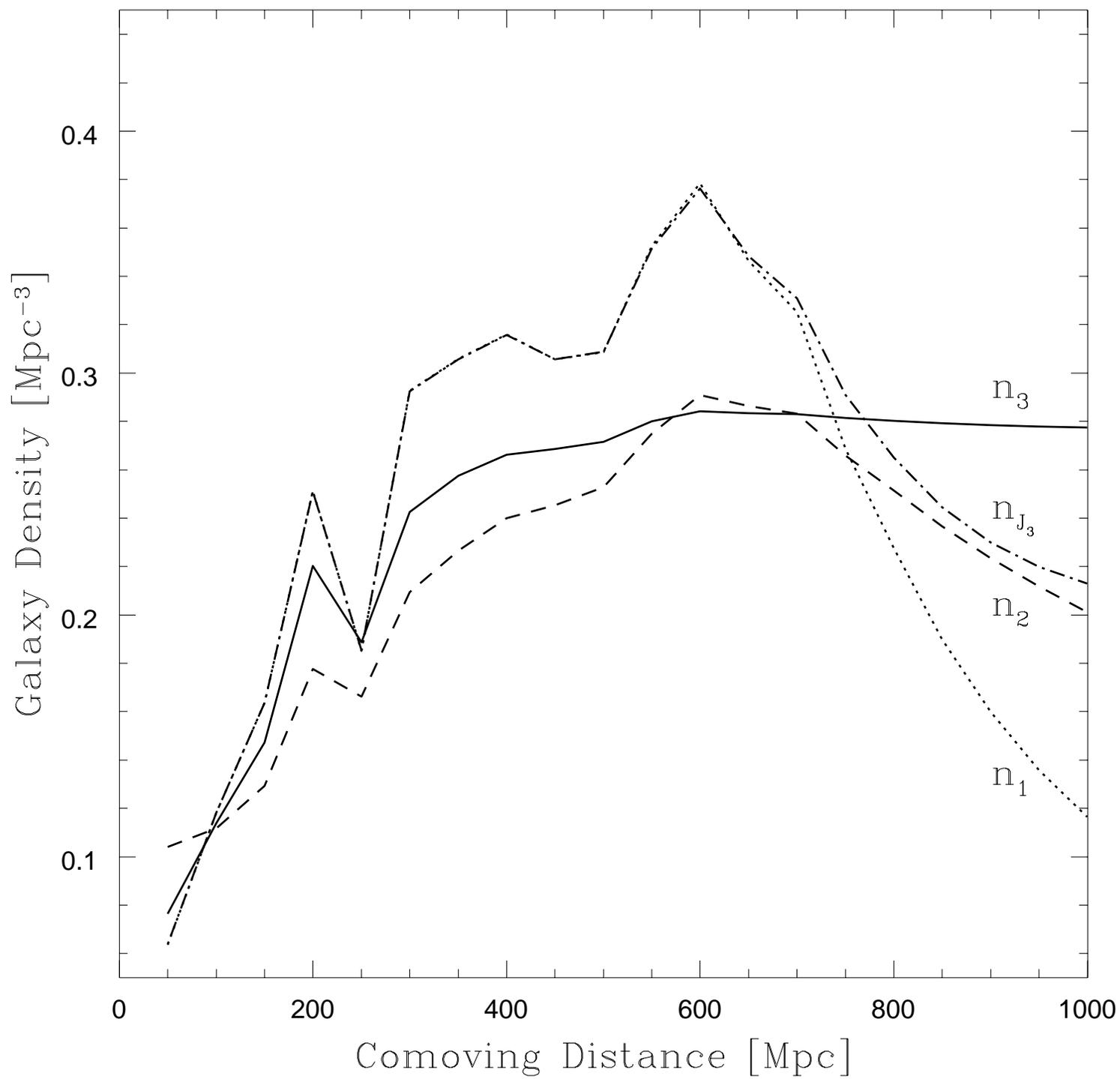
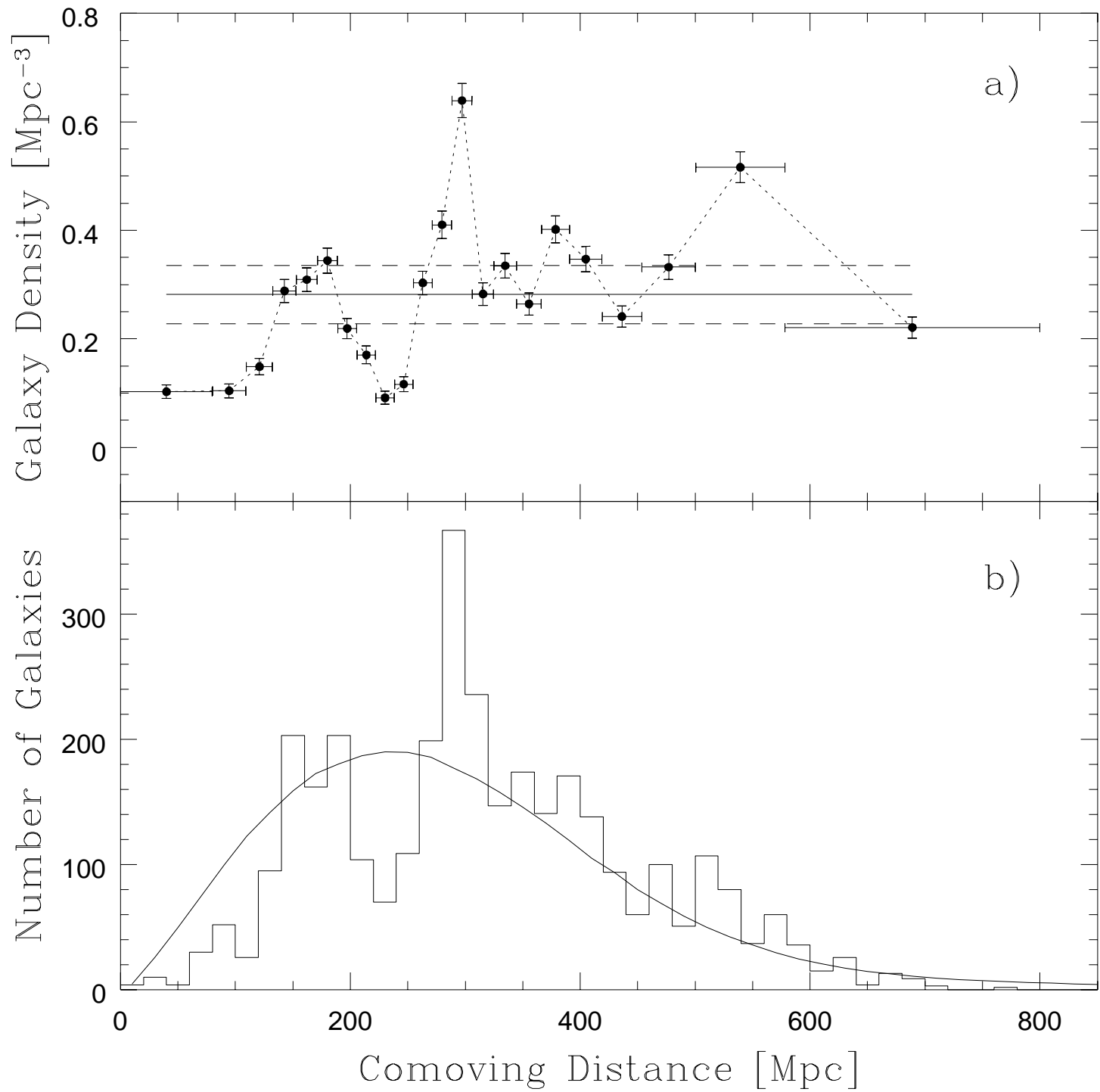
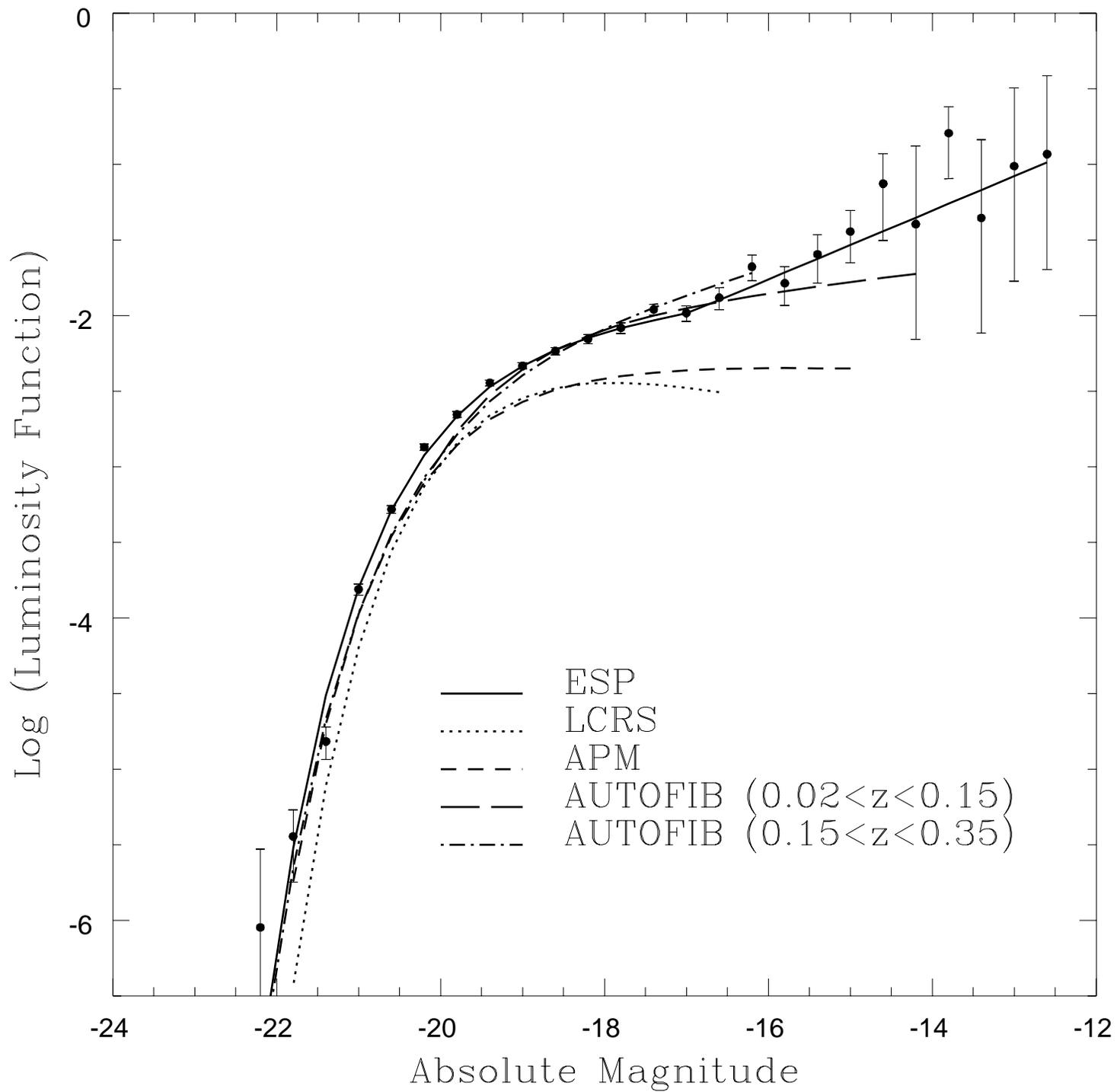
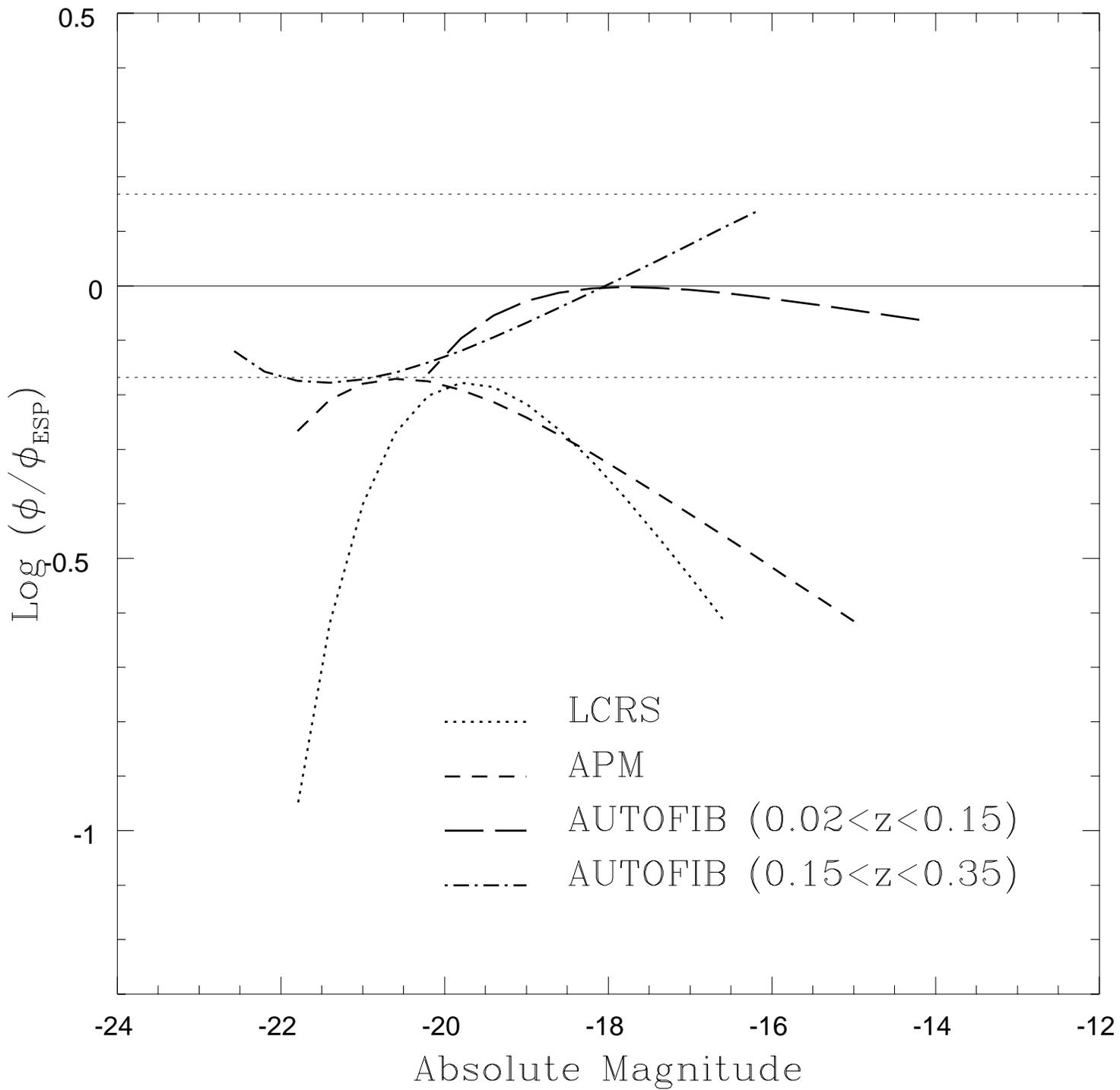
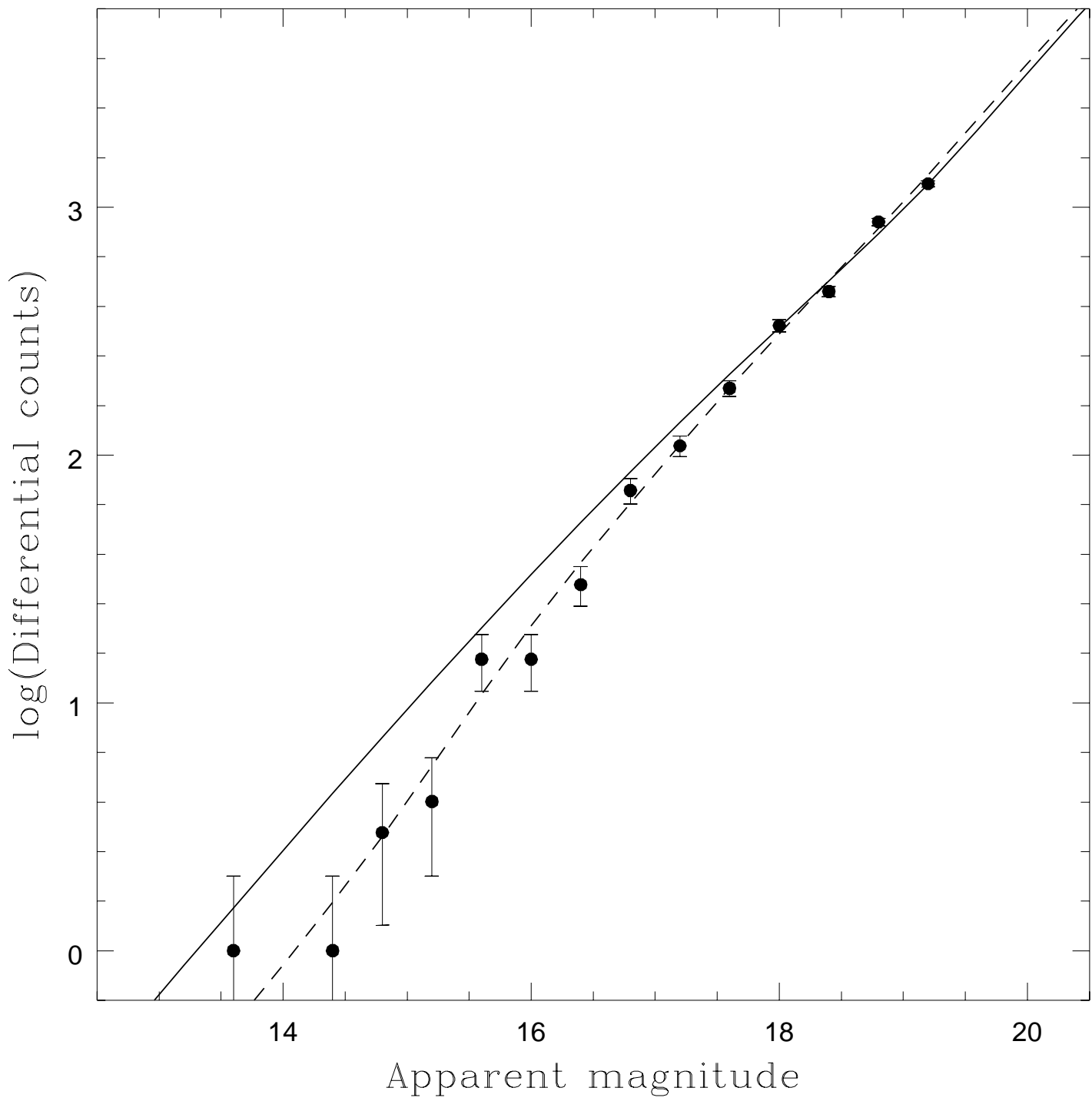
\end{document}